\documentclass[iop,revtex4]{emulateapj}
\usepackage{lineno}

\begin{document}

\renewcommand{\topfraction}{1.0}
\renewcommand{\bottomfraction}{1.0}
\renewcommand{\textfraction}{0.0}

\title{Orbits of four young triple-lined multiple systems}

\author{Andrei Tokovinin}
\affil{Cerro Tololo Inter-American Observatory, Casilla 603, La Serena, Chile}
\email{atokovinin@ctio.noao.edu}

\begin{abstract}
Each  of the nearby  triple systems  HIP 7601,  13498, 23824, and  113597 (HD
10800, 18198, 35877,  217389) consists  of solar-type dwarfs  with comparable
masses, where all three  components are resolved spectrally, while the
outer pairs  are resolved both  visually and spectrally.   These stars
are relatively young (between  100 and 600\,Myr) and chromospherically
active  (X-ray sources), although  they rotate  slowly.  Spectroscopic
orbits of  the inner  subsystems (periods 19.4,  14.1, 5.6, 20.3  days) and
orbits of the outer systems  (periods 1.75, 51, 27, 500 yrs, respectively)
are   determined.    For   HIP    7601   and   13498,   the   combined
spectro-interferometric  outer orbits  produce  direct measurement  of
masses  of all  components, allowing  comparison with  stellar models.
The 6708\AA ~lithium  line is present and its  strength is measured in
each component individually by  subtracting the contributions of other
components.   The  inner and  outer  orbits  of  HIP 7601  are  nearly
circular, likely co-planar,  and have a modest period ratio of 1:33.
This  study  contributes   to  the  characterization  of  hierarchical
multiplicity in  the solar neighborhood and provides  data for testing
stellar evolutionary models and chronology.
\end{abstract}


\maketitle

\section{Introduction}
\label{sec:intro}

Binary   and    multiple   stars   attract    interest   for   various
reasons. Historically, they served as excellent calibrators of stellar
masses, radii,  and other parameters,  e.g.  for testing  and refining
evolutionary models. Multiple  systems also host interesting processes
such    as   dynamical    interactions,   mass    transfer,   mergers,
etc. Multiplicity statistics and studies of certain key objects inform
us on the  processes of star formation.  The  objects featured in this
work are interesting from several points of view.

\begin{deluxetable*}{rr  c l cc rr  c }
\tabletypesize{\scriptsize}     
\tablecaption{Basic parameters of multiple systems
\label{tab:objects} }  
\tablewidth{0pt}                                   
\tablehead{                                                                     
\colhead{HIP} & 
\colhead{HD} & 
\colhead{WDS} & 
\colhead{Spectral} & 
\colhead{$V$} & 
\colhead{$B-V$} & 
\colhead{$\mu^*_\alpha$} & 
\colhead{$\mu_\delta$} & 
\colhead{$\pi_{\rm HIP2}$\tablenotemark{a}} \\
&   &  
\colhead{(J2000)} & 
\colhead{type} & 
\colhead{(mag)} &
\colhead{(mag)} &
\multicolumn{2}{c}{ (mas yr$^{-1}$)} &
\colhead{(mas)} 
}
\startdata
7601  & 10800 & 01379$-$8258 & G1V    & 5.87 & 0.62 & +122 & +120  & 36.52 $\pm$ 0.28 \\
13498 & 18198 & 02539$-$4436 & K0III? & 7.73 & 0.69 & +107 & +56   & 14.09 $\pm$ 0.73 \\
23824 & 35877 & 05073$-$8352 & F8V    & 6.80 & 0.57 & +57  & +147  & 20.81 $\pm$ 0.48 \\
113579& 217343& 23005$-$2619 & G5V   & 7.47 & 0.64 & +7 &$-$108   & 32.51  $\pm$ 0.71  \\ 
113597& 217379& 23005$-$2619 & K7V   & 9.65 & 1.34 & +14&$-$157   & 32.74  $\pm$ 2.03 
\enddata
\tablenotetext{a}{Proper motions and parallax from HIP2 \citep{HIP2}.}
\end{deluxetable*}

Radial  velocities (RVs)  of nearby  solar-type multiple  systems were
monitored in 2014 and 2015 to determine the frequency of spectroscopic
subsystems in  visual components and  to follow their  orbital motion.
The resulting  statistics were  reported in \citep{survey},  while 
detailed  analysis  of  individual  systems  was  deferred  to  future
publications. This paper  is the first of this  series.  It deals with
four  triple-lined spectroscopic  binaries  (SB3s). These  solar-type
multiple  systems (Table~\ref{tab:objects})  are relatively  young, as
inferred from  chromospheric activity (all are X-ray  sources) and the
presence of  lithium. The first  two objects were  identified as
SB3s by \citet{Wichman2003} in a spectroscopic follow-up of {\it ROSAT}. The
wide common-proper-motion (CPM) pair  HIP 113579 and 113597 belongs to
the AB Doradus association  \citep{SACY}; its fainter component is the
SB3. In all four SB3s, the outer subsystems  are spatially resolved;
their orbits  in the  plane of  the sky are  computed or  updated here
together with spectroscopic orbits of the inner subsystems.

In  this  work,  I  designate individual  components  of  hierarchical
multiples by sequences of letters such as A, Ab or AB; a component may
contain one  or several stars.   The binary systems are  designated by
joining components with comma, e.g. Aa,Ab.  The designation A,B stands
for  the  binary  system  with  two  components A  and  B,  while  the
designation AB stands for one  component, e.g. an unresolved binary in a
system AB,C.
 
The  observational  material  and   data  analysis  are  presented  in
Section~\ref{sec:obs}.  Sections 3 to  6 are devoted to the individual
systems  and  follow  the  same template:  bibliography,  orbits,  and
estimation of  component's parameters.  In Section 7,  the results are
used  to discuss  the mass-luminosity relation,  strength of  the lithium
line,  and  origin of  hierarchical  triple-lined  systems.  The  last
Section 8 summarizes the paper.

\section{Observations and data analysis}
\label{sec:obs}

\subsection{Spectroscopic observations}

Most spectra used here were  taken with the 1.5-m telescope located at
the Cerro  Tololo Interamerican Observatory  in Chile and  operated by
the                    SMARTS                    Consortium.\footnote{
  \url{http://www.astro.yale.edu/smarts/}}  The   observing  time  was
allocated through  NOAO (programs 14B-0009,  15A-0055, 15B-0012).  The
observations were  made with the   fiber-fed echelle spectrograph
CHIRON\footnote{\url{http://www.ctio.noao.edu/noao/content/chiron}}
\citep{CHIRON}  by  the  telescope  operators in  service  mode.   The
spectra taken in the slicer mode have a resolution of $R=80,000$ and a
signal to noise  ratio of at least 20.  They  cover the range from
  415\,nm to  880\,nm in  53 orders. Thorium-argon  calibrations were
recorded for  each target.  A  few spectra taken  in 2010 at  the same
telescope  with  the  Fiber   Echelle  (FECH)  with  a  resolution  of
$R=44,000$  \citep{LCO}  are  used  here,  too.  Most  (but  not  all)
individual  RVs  are published  in  \citep{survey}, where  preliminary
orbital periods are also announced.

\subsection{Radial velocities by cross-correlation}

The reduced and wavelength-calibrated  spectra delivered by the CHIRON
pipeline were retrieved from the SMARTS center at the Yale University.
The availability of this service  has greatly enhanced this program by
allowing  rapid analysis  of the  RVs and  flexible scheduling  of new
observations when needed.

\begin{figure}
\plotone{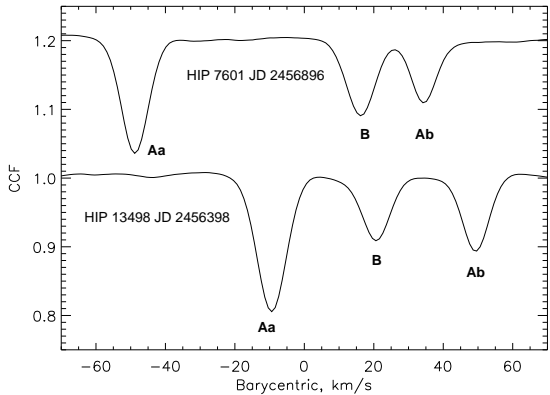}
\caption{CCFs of  triple-lined systems HIP 7601  (upper curve, shifted
  by 0.2)  and HIP  13498 (lower curve).  The letters  mark components
  corresponding to each CCF dip.
\label{fig:CCF}
}
\end{figure}

The  spectra  were  cross-correlated  with  the  digital  binary  mask
(template)  based on  the solar  spectrum stored  in the  NOAO archive
\citep[see][for  more details]{LCO}.   The  cross-correlation function
(CCF)  $C(v)$   is  computed  over   the  RV  span   of  $  v   =  \pm
200$\,km~s$^{-1}$ in the spectral range from 4500\AA ~to 6500\AA. 
  CCFs of all orders are simply summed up and normalized by the median
  value. Portions  of the CCFs within  $\pm 2.35 \sigma$  of each dip
are approximated  by Gaussian curves.  After the  first iteration, the
centers and  dispersions are determined,  and in the  second iteration
the fitting  area is adjusted  accordingly.  The CCF model  with three
Gaussians is
\begin{equation}
C(v) =    1 -   \sum_{j=1}^3 a_j \exp [ -(v - v_j)^2/ 2 \sigma_j^2 ] ;
\label{eq:fit}
\end{equation}
 it contains  nine free parameters $(v_j, a_j,  \sigma_j)$.  CCFs with
 overlapping (blended) dips were processed by fixing amplitudes $a_j$ or
  dispersions  $\sigma_j$ to the values determined  on other nights
 from well-resolved CCFs, like those in Figure~\ref{fig:CCF}.

I  do  not  provide formal  errors  of  RVs  and of  other  parameters
resulting  from the  CCF  fits, as  they  are very  small  and do  not
characterize the real  precision of the results.  The  RV precision is
dominated by systematic effects and is estimated from residuals to the
orbits at  $\sim$0.1 km~s$^{-1}$.  The velocity  zero point relies
  on the wavelength calibration and on the mask derived from the solar
  spectrum, rather than on standard stars.

\subsection{Estimation of rotation  velocity}
\label{sec:vsini}

The CCF width $\sigma$ depends on the spectral resolution, correlation
mask,  and  the intrinsic  width  of  stellar  lines. The  latter  are
broadened  by  rotation, macro-turbulence,  and  other factors.   When
rotation  dominates,  the  CCF  shape deviates  substantially  from  a
Gaussian, but for moderate or slow rotation the CCF is well modeled by
a  Gaussian  with increased  dispersion  $\sigma$.   I calibrated  the
relation between the projected rotation  speed $V \sin i$ and $\sigma$
using  digital solar  spectrum.  It  was broadened  by a  Gaussian PSF
corresponding to the spectral resolution $R=80,000$ (3.75 km~s$^{-1}$)
and   rotation  \citep[e.g.][]{Diaz2011},   assuming  a   linear  limb
darkening law with a coefficient  of 0.65.  The broadened spectrum was
correlated with the  mask. After accounting for the  solar rotation of
2\,km~s$^{-1}$, the results can be fitted by the formula
\begin{equation}
V \sin i \approx 1.80 \sqrt{ \sigma^2 - \sigma_0^2},
\label{eq:Vsini}
\end{equation}
where  $\sigma_0 =  3.4$  and  all values  are  in km~s$^{-1}$.   This
relation is valid for $\sigma <  12$ and is accurate to $\pm$0.4, 
  as long as  line broadening factors other than  rotation are same as
  in the Sun.   In the fiber mode of CHIRON, not used here, $\sigma_0
= 5.4$  and the coefficient  is 1.92.  This approximate  method cannot
replace detailed  analysis of  line profiles, but  it is  suitable for
multiple stars with solar-type components studied here.  Disentangling
of spectra will be needed for a more refined determination of rotation
and other stellar properties. 

\subsection{Speckle interferometry}

Information on the resolved outer subsystems is retrieved from the
Washington Double Star Catalog, WDS \citep{WDS}. It is complemented by
recent speckle interferometry at the SOAR telescope. The latest
publication \citep{SAM15} contains references to the previous papers
of this series. Observations made in 2015 and 2016 are not yet published.  

\subsection{Orbit calculation}

Orbital elements and their errors were determined by the least-squares
fits with  weights inversely proportional to the  adopted data errors.
The          IDL          code          {\tt          orbit}\footnote{
  \url{http://www.ctio.noao.edu/\~{}atokovin/orbit/}} was used. It can
fit   spectroscopic,    visual,   or   combined   visual/spectroscopic
orbits. Formal  errors of orbital  elements are determined  from these
fits.  The errors of masses and orbital parallax derived from combined
orbits  are  computed  by  taking into  account  correlations  between
individual elements.

Orbits of the subsystems were determined iteratively.  Knowing the
semi-amplitudes $K_1$ and  $K_2$ in the inner orbit  Aa,Ab, I estimate
the RV of  A (center of mass)  as the weighted sum, $V_{\rm  A} = (K_2
V_{\rm Aa} + K_1 V_{\rm Ab})/(K_1 + K_2)$. These RVs together with the
RVs of the  secondary component B and the  speckle measurements define
the outer  orbit.  The  $V_{\rm A}$ resulting  from the motion  in the
outer orbit is then subtracted from $ V_{\rm Aa}$ and $ V_{\rm Ab}$ in
the next iteration of the  inner orbit, so its center-of-mass velocity
$\gamma$ is zero by definition.

Spectroscopic  orbital elements  derived in  this work  are  listed in
Table~\ref{tab:sborb},   the   visual    orbits   are   assembled   in
Table~\ref{tab:vborb}, in  common notation.    For circular orbit,
  $\omega=0$ means that the element $T$ corresponds to passage through
  the    node   (maximum    RV).     The   last    two   columns    of
  Table~\ref{tab:sborb}  give  the   weighted  rms  residuals  to  the
  spectroscopic  orbits   and  the  total  number  of   RVs  for  both
  components.  The  last column of  Table~\ref{tab:vborb} contains the
  number of astrometric measures.    The combined orbits are featured
in both tables, duplicating overlapping elements.  In combined orbits,
the longitude of periastron $\omega_A$ corresponds to the primary, and
the  position   angle  of  the  visual  orbit   $\Omega_A$  is  chosen
accordingly to describe  the motion of the secondary.   The weights of
positional measurements  and RVs are  balanced, so that each  data set
has $\chi^2/N  \sim 1$,   where  $N$ is the  number of  degrees of
  freedom.

\begin{deluxetable*}{l l cccc ccc c c}    
\tabletypesize{\scriptsize}     
\tablecaption{Spectroscopic orbits
\label{tab:sborb}          }
\tablewidth{0pt}                                   
\tablehead{                                                                     
\colhead{HIP} & 
\colhead{System} & 
\colhead{$P$} & 
\colhead{$T$} & 
\colhead{$e$} & 
\colhead{$\omega_{\rm A}$ } & 
\colhead{$K_1$} & 
\colhead{$K_2$} & 
\colhead{$\gamma$} & 
\colhead{rms$_{1,2}$} &
\colhead{$N_{1,2}$} 
\\
\colhead{{\it HD}} & & \colhead{(d)} &
\colhead{(+24\,00000)} & &
\colhead{(deg)} & 
\colhead{(km~s$^{-1}$)} &
\colhead{(km~s$^{-1}$)} &
\colhead{(km~s$^{-1}$)} &
\colhead{(km~s$^{-1}$)} &
}
\startdata
7601  &Aa,Ab  & 19.37128        & 56938.599  & 0.1028      & 242.8    & 36.45     & 43.32 & 0.0          & 0.12      & 19 \\
{\it 10800}      &       & $\pm$0.00005 & $\pm$0.021 & $\pm$0.0009 & $\pm$0.4 & $\pm$0.04 & $\pm$0.05 & fixed & 0.15 & 17  \\
7601  & A,B   & 638.67     & 56903.42         & 0.1912      & 151.2 &  9.10 & 17.91 &  $-$1.65 & 0.08               & 19 \\
{\it 10800}  &  & $\pm$0.17  & $\pm$0.94  & $\pm$0.0018 & $\pm$0.3    &  $\pm$0.03 & $\pm$0.05 & $\pm$0.02 & 0.13 & 20  \\
13498 & Aa,Ab &  14.0921   & 56974.21        & 0.328    &  19.3     & 41.96 & 48.66      & 17.78 & 0.06               & 9\\ 
{\it 18198}      &       &  $\pm$0.0025&  $\pm$0.02 & $\pm$0.005& $\pm$1.1& $\pm$0.31 & $\pm$0.34 & $\pm$0.08 & 0.10  & 9\\
13498 & A,B    & 18794   & 54190          & 0.774     & 16.2     & 3.98 & 9.93      & 18.61 & 0.03                    & 10\\ 
{\it 18198}      &     &  $\pm$301&  $\pm$49& $\pm$0.013 & $\pm$1.8 & $\pm$0.61 & $\pm$0.70 & $\pm$0.11 & 0.03        & 9\\
23824 & Aa,Ab  & 5.6310  & 57372.683      & 0 &     0 & 54.72     &  61.64       &   1.29   & 0.12                    & 11 \\
{\it 35877}      &        & $\pm$0.0003&  $\pm$0.003 & fixed &  fixed & $\pm$0.22 & $\pm$0.22 & $\pm$0.12 & 0.16      & 9\\
23824 & A,B   & 9857  & 56783.1      & 0.654     &     70.7     & 4.55     &  11.78       &  4.60   & 0.00            & 11\\
{\it 35877}      &        & $\pm$206&  $\pm$6.6 & $\pm$0.007 & $\pm$0.9    & $\pm$0.29 & $\pm$0.26 & fixed & 0.20     & 9\\
113597& Aa,Ab  & 20.3490         & 57255.941  & 0.433      &  43.4    & 41.06     & 43.62 & 7.56          & 0.55      & 18\\
{\it 217379}      &        & $\pm$0.0002     & $\pm$0.032 & $\pm$0.003  & $\pm$0.6 & $\pm$0.23 & $\pm$0.23 &  $\pm$0.11 & 0.49 & 16 
\enddata 
\end{deluxetable*}

\begin{deluxetable*}{l l cccc ccc  c}    
\tabletypesize{\scriptsize}     
\tablecaption{Visual orbits
\label{tab:vborb}          }
\tablewidth{0pt}                                   
\tablehead{                                                                     
\colhead{HIP} & 
\colhead{System} & 
\colhead{$P$} & 
\colhead{$T$} & 
\colhead{$e$} & 
\colhead{$a$} & 
\colhead{$\Omega_{\rm A}$ } & 
\colhead{$\omega_{\rm A}$ } & 
\colhead{$i$ }  &
\colhead{$N$ }  
\\
\colhead{{\it HD}} & & \colhead{(yr)} &
\colhead{(yr)} & &
\colhead{(arcsec)} & 
\colhead{(deg)} & 
\colhead{(deg)} & 
\colhead{(deg)} &
}
\startdata
7601  & A,B   & 1.74861    & 2014.672       & 0.1912     & 0.07823           & 296.0    & 151.2       &  47.60 & 14\\
{\it 10800} & & $\pm$0.00047  & $\pm$0.003  & $\pm$0.0018 & $\pm$0.00047  & $\pm$0.3   & $\pm$0.3 &  $\pm$0.48 & \\ 
13498 & A,B    & 51.46   & 2007.24     & 0.774   &    0.2967        & 298.3       & 16.2    &  49.5 & 17\\
{\it 18198}      &        &  $\pm$0.82& $\pm$0.14 & $\pm$0.013 &  $\pm$0.0044  & $\pm$1.3   & $\pm$1.7 &  $\pm$1.9 & \\
23824 & A,B    & 26.99   & 2014.34     & 0.654   &    0.2742        & 2.4       & 70.7    &  61.6& 11 \\
{\it 35877}      &        &  $\pm$0.56& $\pm$0.02 & $\pm$0.006 &  $\pm$0.0040  & $\pm$0.7   & $\pm$0.9 &  $\pm$0.50 & \\
113597& A,B   & 500        & 1843.2  & 0.55 &    2.46        & 250.1      & 150.0    &  80.9 & 14
\enddata 
\end{deluxetable*}

The observations used in orbit  calculations are listed in two tables,
published in full  electronically.  Table~\ref{tab:rv} gives, for each
date, the RVs of the primary and secondary components $V_1$ and $V_2$,
their errors used  for relative weighting, and residuals  to the orbit
O$-$C.   The first  column contains  the {\it  Hipparcos}  number, the
second column identifies  the system.  For the inner  subsystem of HIP
7601,   the  velocities   $V_1$  and   $V_2$  are   relative   to  the
center-of-mass $V_{\rm A}$ (motion in the visual orbit is subtracted).
In  the outer  orbits,  $V_1$ refers  to  the center  of  mass of  the
subsystem.  The  last column of Table~\ref{tab:rv}  specifies the data
source.  RVs of the non-variable component HIP 113597B are also listed
in this Table.

The   resolved      astrometric measurements  are   listed   in
Table~\ref{tab:speckle} in  a similar way  as the RVs, with  the first
two columns  identifying the system.   Then follow the  date, position
angle $\theta$,  separation $\rho$, position  error $\sigma_\rho$ used
for weighting, residuals to the orbit, and reference.

\begin{deluxetable*}{r l c rrr rrr l}    
\tabletypesize{\scriptsize}     
\tablecaption{Radial velocities and residuals (fragment)
\label{tab:rv}          }
\tablewidth{0pt}                                   
\tablehead{                                                                     
\colhead{HIP} & 
\colhead{System} & 
\colhead{Date} & 
\colhead{$V_1$} & 
\colhead{$\sigma_1$} & 
\colhead{(O$-$C)$_1$ } &
\colhead{$V_2$} & 
\colhead{$\sigma_2$} & 
\colhead{(O$-$C)$_2$ } &
\colhead{Ref.\tablenotemark{a}} \\ 
 & & 
\colhead{(JD +2400000)} &
\multicolumn{3}{c}{(km s$^{-1}$)}  &
\multicolumn{3}{c}{(km s$^{-1}$)}  &
}
\startdata
7601 & Aa,Ab & 54855.5390 &     20.426 &      0.100 &      0.147 &  $-$24.180 &      0.200 &     $-$0.080 & H \\
7601 & Aa,Ab & 57352.5730 &     31.592 &      0.100 &     $-$0.004 &  $-$37.246 &      0.150 &      0.304 & C\\
7601 & Aa,Ab & 57365.5720 &     $-$7.403 &      0.100 &     $-$0.087 &    8.838 &      0.150 &      0.144 & C\\
 7601 & A,B &  51845.2470 &    $-$12.060 &      2.000 &     $-$0.009 &    18.700 &      1.000 &     $-$0.110 & W \\
 7601 & A,B &  54782.5586 &      6.040 &      1.000 &      0.580 &   $-$15.950 &      1.000 &     $-$0.308 & L \\
 7601 & A,B &  54855.5386 &      2.225 &      0.080 &      0.171 &    $-$9.147 &      0.200 &     $-$0.207 & H \\
 7601 & A,B &  56896.8139 &    $-$10.712 &      0.080 &     $-$0.041 &    16.253 &      0.150 &      0.159 & C 
\enddata 
\tablenotetext{a}{
C: CHIRON;
E: \citet{Elliott2014};
F: FECH;
H: HARPS; 
L: \citet{survey};
W: \citet{Wichman2003}. 
}
\end{deluxetable*}

\begin{deluxetable*}{r l l rrr rr l}    
\tabletypesize{\scriptsize}     
\tablecaption{Position measurements and residuals (fragment)
\label{tab:speckle}          }
\tablewidth{0pt}                                   
\tablehead{                                                                     
\colhead{HIP} & 
\colhead{System} & 
\colhead{Date} & 
\colhead{$\theta$} & 
\colhead{$\rho$} & 
\colhead{$\sigma$} & 
\colhead{(O$-$C)$_\theta$ } & 
\colhead{(O$-$C)$_\rho$ } &
\colhead{Ref.\tablenotemark{a}} \\
 & & 
\colhead{(yr)} &
\colhead{(\degr)} &
\colhead{(\arcsec)} &
\colhead{(\arcsec)} &
\colhead{(\degr)} &
\colhead{(\arcsec)} &
}
\startdata
 7601 & A,B &  2011.0366 &     50.2 &   0.0378 &    0.020 &     $-$3.7 &  $-$0.0096 &SOAR  \\
 7601 & A,B &  2011.0366 &     53.4 &   0.0488 &    0.010 &     $-$0.5 &   0.0014 &SOAR  \\
 7601 & A,B &  2014.7661 &    113.0 &   0.0640 &    0.001 &     $-$3.0 &  $-$0.0004 &SOAR  \\
13498 & A,B  & 1927.0200 &    126.4 &    0.390 &    1.050 &      0.3 &   $-$0.125 & Vis \\
13498 & A,B  & 1990.9160 &    139.2 &    0.444 &    0.005 &      0.1 &    0.004 & Spe  \\
13498 & A,B  & 1991.2500 &    140.0 &    0.437 &    0.010 &      0.5 &    0.002 & Hip 
\enddata 
\tablenotetext{a}{
HIP: Hipparcos;
SOAR: speckle interferometry at SOAR;
Spe: speckle interferometry at other telescopes ;
Vis: visual micrometer measures.
}
\end{deluxetable*}

\subsection{System modeling}

Individual magnitudes of the components are computed from the relative
 areas  of their  CCFs which  are  proportional to  the product  $a
\sigma$.  Although spectral types of all components in SB3s are similar, a
small correction is  needed to account for the  dependence of the line
strength   on   effective   temperature.   Using   synthetic   spectra
\citep{Bertone2008},  I found  that  the CCF  surface  depends on  the
effective  temperature $T_e$  almost  linearly, proportional  to $1  -
3.27(T_e/6000 -1)$ for $4000 <  T_e <6500$. At lower $T_e$ the surface
decreases.  The measured  CCF areas are corrected  for this factor,
assuming a reasonable $T_e$, to split the combined $V$-band flux between
the components.   Resolved photometry  of visual pairs  is used  as an
independent check. The CCF parameters and magnitudes of individual
components are given in Table~\ref{tab:CCF}.

The {\it Hipparcos} distance  modulus gives absolute magnitudes of the
components,  so their  masses are  estimated using  standard relations
\citep[e.g.][]{HM93,FG67a}       or       Dartmouthh       isochrones
\citep{Dotter2008}, assuming that the  stars are on the main sequence.
Effective temperatures and  individual magnitudes in different filters
are then  derived from the masses  with the help of  the same standard
relations, and the combined photometry resulting from the system model
is compared to  the actual combined magnitudes. This  provides a check
on the temperatures. The modeled  masses are compared to the dynamical
mass sum from the visual orbit or to the actually measured masses from
the combined orbit.  Adjustments to  the model are made iteratively to
achieve a better match between observed and modeled properties of each
system; the details are different, depending on the available data.

The  guesswork of  matching  data and    models cannot  replace
actual measurement  of stellar parameters.  However, it  is useful for
detecting  discrepancies and  for estimating  parameters  not measured
directly,  such as  effective  temperatures and  colors of  individual
components.   To distinguish  model-dependent  (estimated) parameters,
they are marked by asterisks in the tables.

\subsection{The lithium line}

Presence  of  the  resonance   lithium  doublet  at  6707.761\AA  ~and
6707.912\AA  ~(average  wavelength   6707.82\AA)  in  the  spectra  of
solar-type   stars  is   a  sign   of  their   relatively   young  age
\citep{Soderblom1993}.  Unfortunately,  this line is  located near the
end of the CHIRON echelle order.  Its quantitative analysis in SB3s is
complicated because the line is a blend of three components in orbital
motion. The following procedure was used. 

The echelle  order containing the  lithium line was normalized  by the
fitted   continuum.   Nominal  positions   of  the   6707.82\AA  ~line
corresponding to each component were computed using their measured RVs
and  the  barycentric  correction.    The  expected  profiles  of  the
lithium-line  blend   computed  with  preliminary   estimates  of  its
amplitudes  and  widths  for  each component  matched  the  individual
spectra quite  well.  Then  I reconstructed the  line profile  of each
component by subtracting the contributions of the other two components
from each spectrum, shifting  the resulting spectrum to the rest-frame
wavelength, and averaging all spectra with weights proportional to the
flux.   The  primary component  with  the  strongest  lithium line  is
processed  first,  using  preliminary  estimates  for  the  other  two
components.   Then  the  line  profile  of  the  second  component  is
reconstructed  in the  same way,  using  the refined  estimate of  the
primary.  Finally,  the weakest component is  retrieved by subtracting
the first two.

The lithium line in the  co-added spectrum of each component is fitted
by  a Gaussian  curve to  measure  its amplitude  $a_{\rm Li}$,  width
$\sigma_{\rm  Li}$,  equivalent width  EW$_{\rm  Li}$, and  wavelength
$\lambda_{\rm Li}$.  The latter  shows that the  lines are  indeed not
shifted  from the  assumed  rest-frame wavelength  of 6707.82\AA.  The
measured EWs are divided by  the relative flux of each component $f_R$
to make  the result comparable to  the line strength  in single stars.
The $R$-band  magnitudes derived from  the system models are  used for
this correction.  Note that  the lithium line  is wider than  the CCF,
being a blend of two lines spaced by 6.7\,km~s$^{-1}$.

\section{HIP 7601 (HD 10800)}
\label{sec:7601}

The G1V star HIP 7601 (HR 512) is located at a distance of 27\,pc from
the Sun; it  is designated as GJ 67.1 in the  catalog of nearby stars.
According  to \citet{Wichman2003},  it is  a young  SB3 system  and an
X-ray  source.  \citet{N04}  recognized it  as a  double-lined binary.
This  bright  star has  an  extensive  literature.  Its  chromospheric
activity was measured by \citet{Henry1996}, $\log R'(HK)= -4.6$. These
authors noted that  the star is slightly more  active than normal, but
not {\it very} active. There is no obvious emission in the core of the
H$\alpha$ line.  \citet{SACY} detected the lithium line 6707.8\AA ~and
measured its equivalent width of  70\,m\AA, indicative of a young age.
Based  on  the  {\it  IRAS}   photometry,  a  mid-IR  excess    at
  25\,$\mu$m  was  detected,  suggesting  a  debris  disk.   However,
further  observations  at the  VLT  have  shown  no mid-IR  excess  at
11.59\,$\mu$m  and   18.72\,$\mu$m  \citep{Smith2008}.   Incidentally,
there  were no  other faint  stars within  the $19''$  field  of view.
\citet{Trilling2008} also found no  energy excess in the {\it Spitzer}
passbands; the flux at 70\,$\mu$m  was even less than predicted by the
photospheric model.

Physical  parameters of  the  star were  derived from  high-resolution
spectra by several  authors, apparently disregarding its multiplicity.
\citet{Maldonado2012}  determined the  metallicity  [Fe/H]=$-0.11$ and
age $\log  t =  9.09$.  \citet{Ammler2012} found  projected rotational
velocity $V \sin i = 1.8 \pm 0.6$ km~s$^{-1}$.  \citet{Casagrande2011}
give  $T_e  =  5905$\,K,   [Fe/H]=$-0.13$,  $\log  g  =  4.06$,  while
\citet{Saffe2008}  determined  $T_e  =  5901$\,K,  $\log  g  =  4.84$,
      [Fe/H]=+0.09.   The triple-lined  nature of  the  spectrum could
      potentially bias these results.

\begin{deluxetable}{r l cccc}    
\tabletypesize{\scriptsize}     
\tablecaption{CCF parameters and $V$ magnitudes
\label{tab:CCF}          }
\tablewidth{0pt}                                   
\tablehead{                                                                     
\colhead{HIP} & 
\colhead{Comp.} & 
\colhead{$a$} & 
\colhead{$\sigma$} & 
\colhead{$a \sigma$} & 
\colhead{$V$} \\
&    &   & (km~s$^{-1}$) &  (km~s$^{-1}$) & (mag) 
}
\startdata
7601 & Aa   & 0.164 & 3.79 &  0.620 & 6.64 \\
7601 & B    & 0.115 & 4.16 &  0.477 & 7.08 \\
7601 & Ab   & 0.086 & 3.56 &  0.307 & 7.74 \\
13498& Aa   & 0.192 & 4.15 &  0.795 & 8.31 \\ 
13498& Ab   & 0.106 & 3.75 &  0.397 & 9.36 \\
13498& B    & 0.092 & 3.93 &  0.360 & 9.52 \\
23824& Aa   & 0.133 & 5.21 &  0.690 & 7.30 \\   
23824& Ab   & 0.089 & 4.61 &  0.411 & 8.13 \\   
23824& B    & 0.042 & 3.52 &  0.151 & 9.61 \\   
113597 & Aa & 0.212 & 5.24 & 1.111  & 10.64 \\
113597 & Ab & 0.130 & 5.57 & 0.724  & 10.80 \\
113597 & B  & 0.081 & 5.96 & 0.483  & 11.09
\enddata 
\end{deluxetable}   

All  three dips in  the CCF  are narrow,  differing in  their contrast
(Figure~\ref{fig:CCF}).   The   dip  amplitudes  of  Aa,   B,  and  Ab
(strongest  to  weakest)  and  the  widths  of  their  CCFs  $\sigma$,
determined  only  from  the  triple-lined  (not blended)  spectra  and
averaged, are  listed in Table~\ref{tab:CCF}.  The $\sigma$ correspond
to  $V  \sin  i$  of  3.0,  4.3,  and 1.8  km~s$^{-1}$  in  Aa,  B,  and  Ab,
respectively (see Section~\ref{sec:vsini}).

The outer  system was  resolved by speckle  interferometry at  SOAR in
2014   and  got   the   WDS  designation   J01379$-$8259  or   TOK~426
\citep{SAM15}. Archival speckle observation made in 2011, distorted by
telescope  vibrations, was  re-processed and  used here  in  the orbit
calculation, as  well as the recent yet  unpublished speckle measures.
The median  magnitude difference is 0.79  mag at both  540 and 770\,nm
wavelengths.  The relative photometry is based on 4 and 5 measurements
in  each band,  with  the rms  scatters  of 0.12  and  0.09 mag.   The
$V$-magnitudes in Table~\ref{tab:CCF} correspond to $\Delta V_{\rm AB}
= 0.77$ mag, in agreement with the speckle photometry. 

\begin{figure}
\plotone{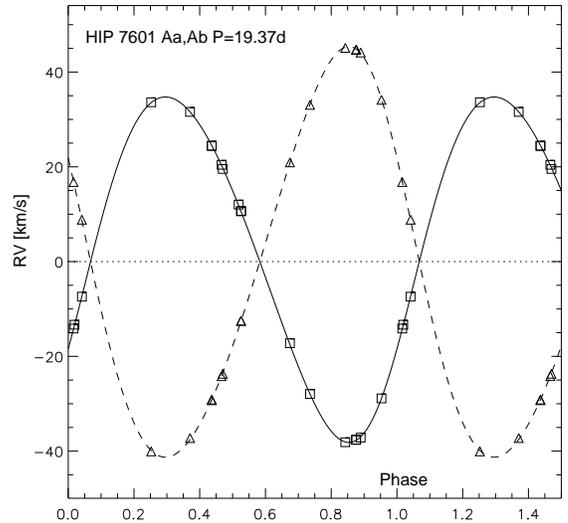}
\caption{ Orbit  of the inner  subsystem HIP 7601 Aa,Ab.   Squares and
  triangles  denote  RVs of  the  primary  Aa  and the  secondary  Ab,
  respectively. Motion in the outer orbit has been subtracted.
\label{fig:7601in}
}
\end{figure}

\begin{figure}
\plotone{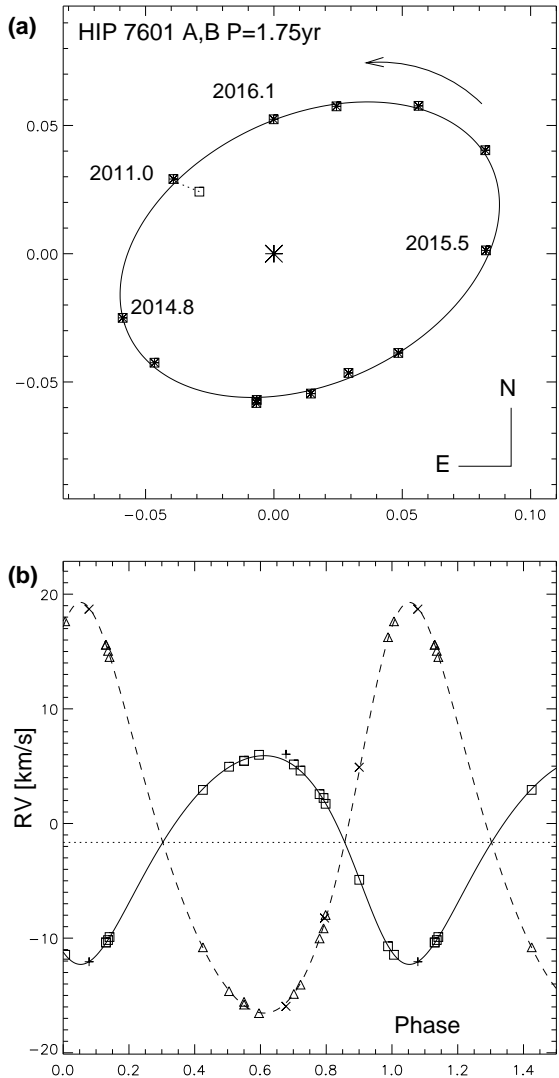}
\caption{Combined orbit  of the outer  system HIP 7601 A,B.   Top (a):
  orbit  in the  plane of  the sky,  with primary  component A  at the
  coordinate origin and scale in arcseconds.  Bottom (b): the RV curve
  (squares for the  center-of-mass A, triangles for B).  RVs  with low
  weight are plotted as crosses.
\label{fig:7601out}
}
\end{figure}

I found in the ESO archive  11 spectra of HIP~7601 taken with HARPS in
the   period   from  2009   to   2012   (program   088C.513  and   its
continuation). The spectral resolution is very similar to CHIRON, $R =
80,000$.   The reduced  spectra were  retrieved from  the  archive and
cross-correlated with the solar mask  to derive the RVs.  Use of these
additional  data does  not increase  the  time base,  but reduces  the
errors  of   the  orbital  elements.    I  applied  a   correction  of
$-0.25$\,km~s$^{-1}$    to   these    RVs,    determined   from    the
$\gamma$-velocity of Aa,Ab fitted to the HARPS data alone. RVs derived
from blended CCFs are not used in the determination of the inner orbit
and are given a low  weight  (large errors in Table~\ref{tab:rv})
in the outer  orbit.  The adopted errors of the  RVs match the actual
rms residuals, yielding $\chi^2/N \sim 1$.

The   orbit    of   the   inner   subsystem   Aa,Ab    is   shown   in
Figure~\ref{fig:7601in},  and   the  combined  spectro-interferometric
orbit of  the outer pair A,B is  featured in Figure~\ref{fig:7601out}.
In the final  iteration of the outer orbit, I added  with a low weight
the  RVs  from \citet{Wichman2003}  measured  in  2000, $-5.4,  -20.6,
+18.4$  km~s$^{-1}$ for  Aa, Ab,  and B,  respectively.   The spectrum
taken in 2008 \citep{LCO} was also used, although it shows only double
lines (two  components were  blended). The rms  residuals to  the SOAR
speckle measures are  1\fdg0 in angle and 1.2\,mas  in separation. The
1.75-yr outer  period is determined  with a relative error  of 0.027
per cent, or just 4 hours.

The combined orbit of A,B leads  to the orbital parallax of $37.10 \pm
0.45$\,mas, in  good agreement  with the HIP2  parallax of  $36.52 \pm
0.28$\,mas.  The  HIP2 parallax  corresponds to the  mass sum of  3.25 ${\cal
  M}_\odot$, while the orbital masses of A and B are $2.033 \pm 0.049$
and  $1.033  \pm  0.026$  ${\cal  M}_\odot$.   Using  the  mass  ratio
$q=0.841$ in the  inner orbit, the masses of Aa  and Ab are determined
as  $1.104  \pm  0.027$   and  $0.929  \pm  0.022$  ${\cal  M}_\odot$.
Uncertainty of  the orbital inclination $i_{\rm A,B}$ contributes  to the errors
of orbital masses and parallax.

By comparing the sum  of orbital masses of  Aa and Ab with $M  \sin^3 i =
1.00$  ${\cal M}_\odot$ in  the inner  orbit, its  inclination $i_{\rm
  Aa,Ab} = $51\fdg9  is derived.  It is similar  to the inclination of
47\fdg4 in  the outer  orbit; the  two orbits thus  could be  close to
co-planarity. The ratio of periods is 32.970$\pm$0.009. Its difference
from the  integer number 33  is small, although  formally significant.
An almost integer period ratio and orbits with low eccentricities that
are  likely  co-planar  make  this  system  similar  to  HD~91962,  the
``planetary'' quadruple with a period ratio of 19.0 \citep{Planetary}.

\begin{deluxetable}{l c c c }    
\tabletypesize{\scriptsize}     
\tablecaption{Components of  HIP 7601
\label{tab:7601}          }
\tablewidth{0pt}                                   
\tablehead{                                                                     
\colhead{Parameter} & 
\colhead{Aa} & 
\colhead{B} & 
\colhead{Ab} 
}
\startdata
Mass (${\cal M}_\odot$)   & 1.10 & 1.03 & 0.93 \\
                         & $\pm$0.03 & $\pm$0.03 & $\pm$0.02 \\
$M_V$ (mag)       &  4.45 & 4.89 & 5.55 \\
$T_e^*$ (K)       & 6131 & 5879 & 5524 \\
$(B-V)^*$ (mag)     & 0.50 & 0.64 & 0.79 \\ 
$\lambda_{\rm Li}$ (\AA) & 6707.82 & 6707.82 & 6707.83 \\
$a_{\rm Li}$ & 0.137    & 0.058  & 0.019 \\
              & $\pm$0.003 & $\pm$0.002 & $\pm$0.002 \\
$\sigma_{\rm Li}$ (km~s$^{-1}$) & 5.25 & 5.71 & 4.23 \\
                   & $\pm$0.13 & $\pm$0.25 & $\pm$0.58 \\
$EW_{\rm Li}$ (m\AA)       & 40.3 & 18.6 & 4.5 \\
$f_R^*$       &  0.46 & 0.33 & 0.21 \\ 
$EW_{\rm Li corr}$ (m\AA)  & 91.3  & 56.7 & 20.6 
\enddata 
\end{deluxetable}   

\begin{figure}
\plotone{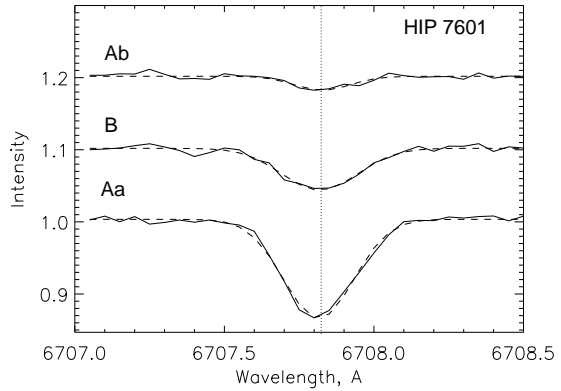}
\caption{Average spectrum in  the region of the lithium  line for each
  component  of HIP 7601,  with the  other two  components subtracted.
  The  curves   (top  to  bottom,  shifted   vertically  for  clarity)
  correspond to  Ab, B, and Aa.   Full lines show  the spectra, dashed
  lines are Gaussian fits. Statistical errors are $\pm$0.005. 
\label{fig:Li7601}
}
\end{figure}

The masses  and $V$-magnitudes  of all three  stars are  measured, the
distance is  known, so the  absolute magnitudes $M_V$ are known,  too.  They
match well  empirical relations between mass and  absolute magnitude  $M_V$, while
the Dartmouth  isochrones predict $M_V$ that are 0.2  mag too
bright  (see Section~\ref{sec:mass}).  To evaluate  effective temperatures
and  colors,  I  interpolated  the solar-metallicity  isochrone  using
masses  reduced  by  5\%.   The  resulting parameters  are  listed  in
Table~\ref{tab:7601},  marked  by  asterisks  to show  that  they  are
guessed rather  than measured.   They lead to  the combined  $B-V$ and
$V-K_s$  colors of  0.60 and  1.44 mag,  respectively, while  the actual
colors are  0.62 and 1.47 mag.   The flux-weighted $T_e  = 5939$\,K is
close to $T_e  \approx 5900$\,K reported in the  literature.  This agreement
indicates  that the  guessed temperatures  of the  stars are  close to
reality.

The  profiles  of  the  lithium  line  in  the  individual  components
(Figure~\ref{fig:Li7601}) are  symmetric (no apparent  blends) and not
shifted  in wavelength with  respect to  their expected  position. The
nine individual  spectra summed in  the combined profile have  a total
flux of 45\,000 electrons per pixel, hence the signal-to-noise ratio (SNR) of
210.   The  sum of  the  EWs is  63.4\,m\AA,  in  good agreement  with
70\,m\AA   ~measured   by  \citet{SACY}.    In   the   last  line   of
Table~\ref{tab:7601} the measured EWs are divided by the relative flux
of  each component  $f_R$, making  the result  directly  comparable to
single stars.   The component  Ab is measured  with the  largest error
owing  to  the small  amplitude  of its  lithium  line  and the  large
correction. 

The    spatial   velocity   of    HIP~7601   calculated    using   the
$\gamma$-velocity    of   A,B    is   $(U,W,V)    =    (-20.0,   -2.4,
  -9.6)$\,km~s$^{-1}$  (the $U$   axis  points  away   from  the  Galactic
center). It corresponds to the  kinematics of young disk. The velocity
is accurate to better than  1\,km~s$^{-1}$. The object does not belong
to any  known kinematic group.   The closest kinematic neighbor  is probably
the Castor  group with  $(U,W,V) = (-10.7,  -7.5, -8.8)$\,km~s$^{-1}$.
According  to \citet{Klutch2014},  the probability  of  the HIP~7601's
membership in the  Castor group could be around  20\%, considering the
velocity dispersion of its known members.

\section{HIP 13498 (HD 18198)}
\label{sec:13498}

The  star  has  a   spectral  type  G8/K1III+F:  in  SIMBAD
\citep{Houk1978},   while     in   fact  it   consists  of   three
  main-sequence dwarfs.  Its color is $B-V=0.69$ mag.  The HIP2 parallax
is    14.09$\pm$0.73\,mas,    the    original   HIP1    parallax    is
14.87$\pm$1.13\,mas.  It is a visual binary I~1480 (WDS J02539$-$4436)
with an  orbital period of  52\,yr, semi-major axis of  0\farcs27, and
high  eccentricity  $e=0.83$  \citep{Hrt2012a}.   There are  no  other
visual components  in the WDS.   \citet{Wichman2003} discovered double
lines  and concluded that  there is  a spectroscopic  subsystem.  They
measured the lithium line with  an equivalent width of 110\,m\AA ~(sum
of  both  components)  and  computed  the relative  X-ray  flux  $\log
L_X/L_{\rm  bol}   =  -4.76$.   Double   lines  were  also   noted  by
\citet{N04}.   \cite{LCO}  detected triple  lines,  implying that  all
three   components  have   comparable   luminosities.   The   physical
parameters  were  derived  from  high-resolution  spectra,  apparently
disregarding the multiplicity, by \citet{Casagrande2011} who give $T_e
= 5925$\,K, [Fe/H]=$0.14$, $\log g = 4.03$.

Relative photometry of the  outer system A,B by speckle interferometry
at SOAR gives the median magnitude differences of 1.52 and 1.49 mag at
540 and  770\,nm, respectively. \citet{Wichman2003}  list the resolved
{\it  Tycho} photometry  of  A and  B  reduced to  the standard  $B,V$
system:  $V = (7.97,  9.54)$ mag  and $B-V=  (0.67, 0.79)$  mag, while
$\Delta Hp_{\rm AB}  = 1.58$\,mag.  The resolved photometry  indicates that the
component  B  is  slightly  redder  than  A.   The  $V$-magnitudes  in
Table~\ref{tab:CCF} correspond  to $\Delta V_{\rm AB} =  1.53$ mag.

\begin{figure}
\plotone{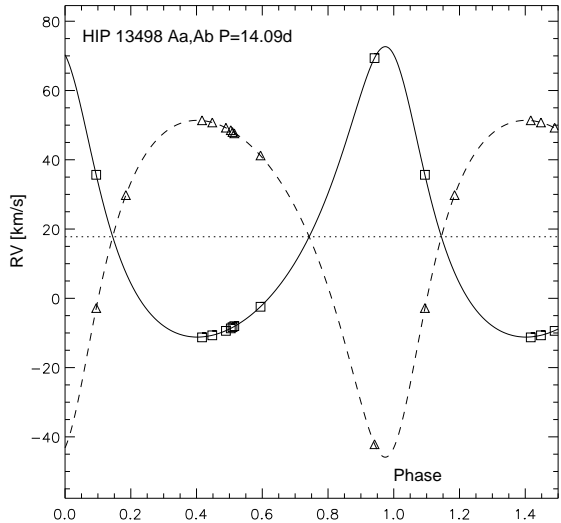}
\caption{Spectroscopic orbit of HIP 13498 Aa,Ab, $P=14.09$ d.
\label{fig:13498a}
}
\end{figure}

\begin{figure}
\plotone{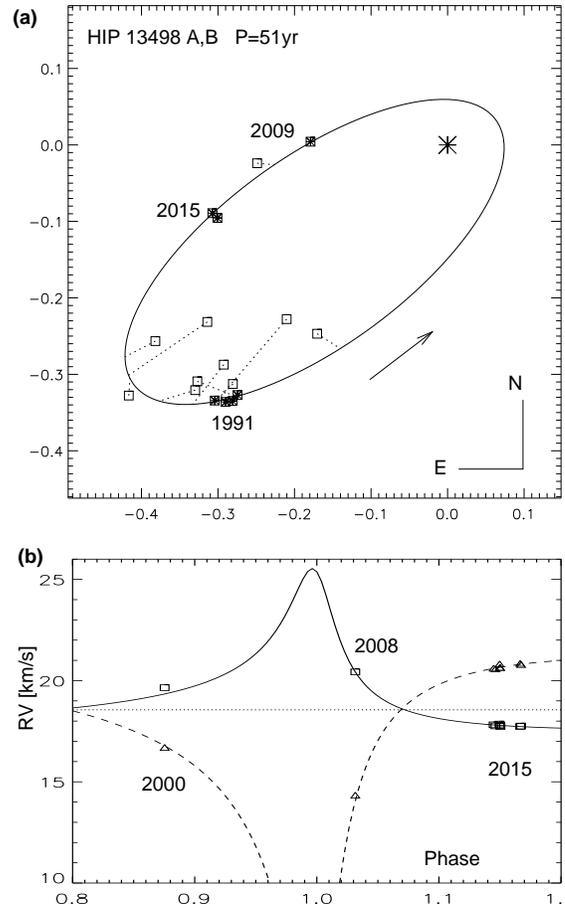}
\caption{Combined  orbit  of  HIP  13498  A,B  (I  1480),  $P=51$\,yr.
  Positions on  the sky (a, top)  have scale in  arcseconds; the empty
  squares denote visual measurements, the filled squares are accurate
  data from speckle and {\it  Hipparcos}.  Fragment of the RV curve is
  shown at the bottom (b).
\label{fig:I1480}
}
\end{figure}

After the initial orbital solutions were derived from the CHIRON data,
I included the RVs measured  in 2008 \citep{LCO}, with a correction to
account  for the  visual orbit  and a  correction of  $+0.5$  km~s$^{-1}$ for
the instrument zero point.   The 14-day orbit of Aa,Ab  is determined quite
reliably, with the  rms RV residuals of  60 and 100 m~s$^{-1}$ for  Aa and Ab,
respectively.

Using the available  micrometer and speckle measurements, the
RVs of  A (center-of-mass  of Aa,Ab) and  B, I   computed  the combined
visual/spectroscopic orbit of  A,B (Figure~\ref{fig:I1480}).  The node
of the  visual orbit  is now defined.   The weighted rms  residuals in
angle and  separation are 0\fdg6  and 3\,mas, respectively,  while the
weighted RV  residuals for A  and B are  35 and 33 m~s$^{-1}$.   The spectrum
obtained  by \citet{Wichman2003}  on 2000.829  (JD 2451847.307)  has a
strong line with $RV= 12.9$  km~s$^{-1}$ (apparently a blend of Aa and
B)  and a  weaker line  at 31.2\,km~s$^{-1}$  that corresponds  to Ab.
Owing to  unresolved blending, this  observation is of little help for
orbit refinement; it is used with a low weight for the orbit of A,B.

The periastron of the A,B orbit  in 2007.3 has not been covered either
by speckle observations (back  then, time allocation committees denied
access  of  speckle  instruments  to  4-m  telescopes)  or  by  radial
velocities (I have not found any spectra in the ESO archive). The next
periastron passage  will happen in  2058,  when the  RV difference
  between A and B will  reach 23 km~s$^{-1}$.  Available data loosely
define the RV  amplitudes because the shape of the  RV curve is poorly
constrained.  Decreasing the element $\omega_A$ by only 2\degr 
  ~increases  the   mass sum by $0.6\,{\cal M}_\odot$.
The unconstrained orbit  in Tables~\ref{tab:sborb} and \ref{tab:vborb}
yields the  orbital parallax $\pi_{\rm  orb} = 14.8 \pm  0.8$\,mas and
the masses $M_A = 2.16 \pm  0.40$ ${\cal M}_\odot$ and $M_B = 0.87 \pm
0.25$ ${\cal M}_\odot$.   This is close to the  HIP1 parallax, which I
adopt here.  The HIP2 parallax leads to the dynamical mass sum of $3.5
{\cal M}_\odot$ that  appears too large, given the  combined color and
magnitude.

Considering  the  uncertainty of  the  orbital  masses,  I adopt  mass
estimates  that are  compatible with  the measurements  and  match the
expected relation between mass  and absolute magnitude.  Starting with
the mass sum of 3.0  ${\cal M}_\odot$ and the well-measured mass ratio
of 0.862 in the  inner orbit, I fix the mass ratio  in the outer orbit
at   $q_{\rm   AB}=0.45$  (compatible   to   its  measurement   within
the uncertainty)   and   obtain   the   component's   masses   listed   in
Table~\ref{tab:13498}.  A  smaller value  of $q_{\rm AB}$  would imply
that the component B is too bright for its mass.

The  effective temperatures  are  estimated from  the Dartmouth  1-Gyr
isochrone  using the  above  masses  reduced by  5\%.  They match  the
combined color of the system and the ``blended'' $T_e = 5925$\,K given
by \citet{Casagrande2011}.  The modeled  combined color $V-K_s = 1.48$
mag agrees  with the measured $V-K_s  = 1.49$ mag;  the model predicts
$B-V  = 0.63$  mag, slightly  bluer than  the measured  0.69  mag.  Of
course, the adopted  $T_e$ remain only a plausible  guess until a more
detailed  analysis is  done  to separate  the  individual spectra  and
actually measure the $T_e$.

\begin{figure}
\plotone{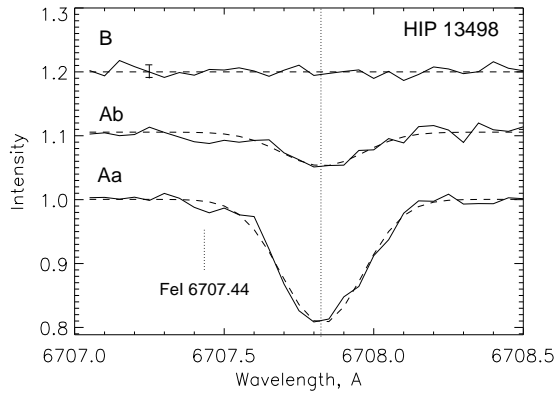}
\caption{Average  spectrum  in the  region  of  the  lithium line  for
  components of  HIP~13498. The full  lines show the data,  the dashed
  lines are Gaussian fits.
\label{fig:13498Li}
}
\end{figure}

The lithium  line is  detected in  Aa and Ab,  and is  below the detection
threshold  in B.  The line  profiles (with  the contribution  of other
components  subtracted) are  given in  Figure~\ref{fig:13498Li} (total
count 17\,000  electrons, SNR=130). A  weak Fe\,I line at  6707.44\AA ~is
noticeable, as well as some asymmetry of the lithium line itself.  The
parameters  of the Gaussian  fits to  the lithium  line are  listed in
Table~\ref{tab:13498}.   

\begin{deluxetable}{l c c c }    
\tabletypesize{\scriptsize}     
\tablecaption{Components  of  HIP 13498
\label{tab:13498}          }
\tablewidth{0pt}                                   
\tablehead{                                                                     
\colhead{Parameter} & 
\colhead{Aa} & 
\colhead{Ab} & 
\colhead{B} 
}
\startdata
Mass$^*$ (${\cal M}_\odot$)   & 1.11 & 0.96 & 0.93 \\
$M_V$ (mag)       &  4.17 & 5.22 & 5.38 \\
$T_e^*$ (K)       & 6103 & 5558 & 5443 \\
$(B-V)^*$ (mag)     & 0.51 & 0.78 & 0.82 \\ 
$\lambda_{\rm Li}$ (\AA) & 6707.83 & 6707.83 & \ldots \\
$a_{\rm Li}$ & 0.193    & 0.053  & $<$0.01 \\
              & $\pm$0.005 & $\pm$0.005 &   \\
$\sigma_{\rm Li}$ (km~s$^{-1}$) & 6.02 & 6.05 & \ldots \\
                   & $\pm$0.17 & $\pm$0.70 &  \\
$EW_{\rm Li}$ (m\AA)       & 65 & 18 & $<$2 \\
$f_R^*$      &  0.52 & 0.26 & 0.22 \\ 
$EW_{\rm  corr}$ (m\AA)  & 125  & 69 & $<$9 
\enddata 
\end{deluxetable}   

The inner orbit corresponds to the mass sum $M \sin^3 i = 0.91$ ${\cal
  M}_\odot$.   The  adopted masses  imply  an  orbital inclination  of
49\degr,  similar  to the  51\fdg5  inclination  of the  outer
orbit. The two orbits could  be nearly co-planar, although this cannot
be established  without resolving the inner  subsystem and determining
the orientation of its orbit in the sky.

The   spatial    velocity   of   HIP~13498    calculated   using   the
$\gamma$-velocity    of   A,B   is    $(U,W,V)   =    (-37.5,   -21.8,
-4.3)$\,km~s$^{-1}$,  placing the  object in  the Hyades  moving group
with    $(U,W,V)   =    (-39.7,   -17.7,    -2.4)$\,km~s$^{-1}$,   see
\citet{Klutch2014}.

\section{HIP 23824 (HD 35877)}
\label{sec:23824}

The star  has a  spectral type  F8V, $B-V=0.57$ mag.   It is  a visual
binary HDS~669 with an orbital period of 26.5\,yr \citep{SAM15}.  
Relative photometry of the  outer system A,B by speckle interferometry
at SOAR gives the median magnitude differences of 2.65 and 2.20 mag at
540 and 770\,nm, respectively, while  $\Delta Hp = 2.53 \pm 0.25$ mag.
The $V$-magnitudes in Table~\ref{tab:CCF} correspond to $\Delta V_{\rm
  AB} = 2.72$ mag.

\begin{figure}
\plotone{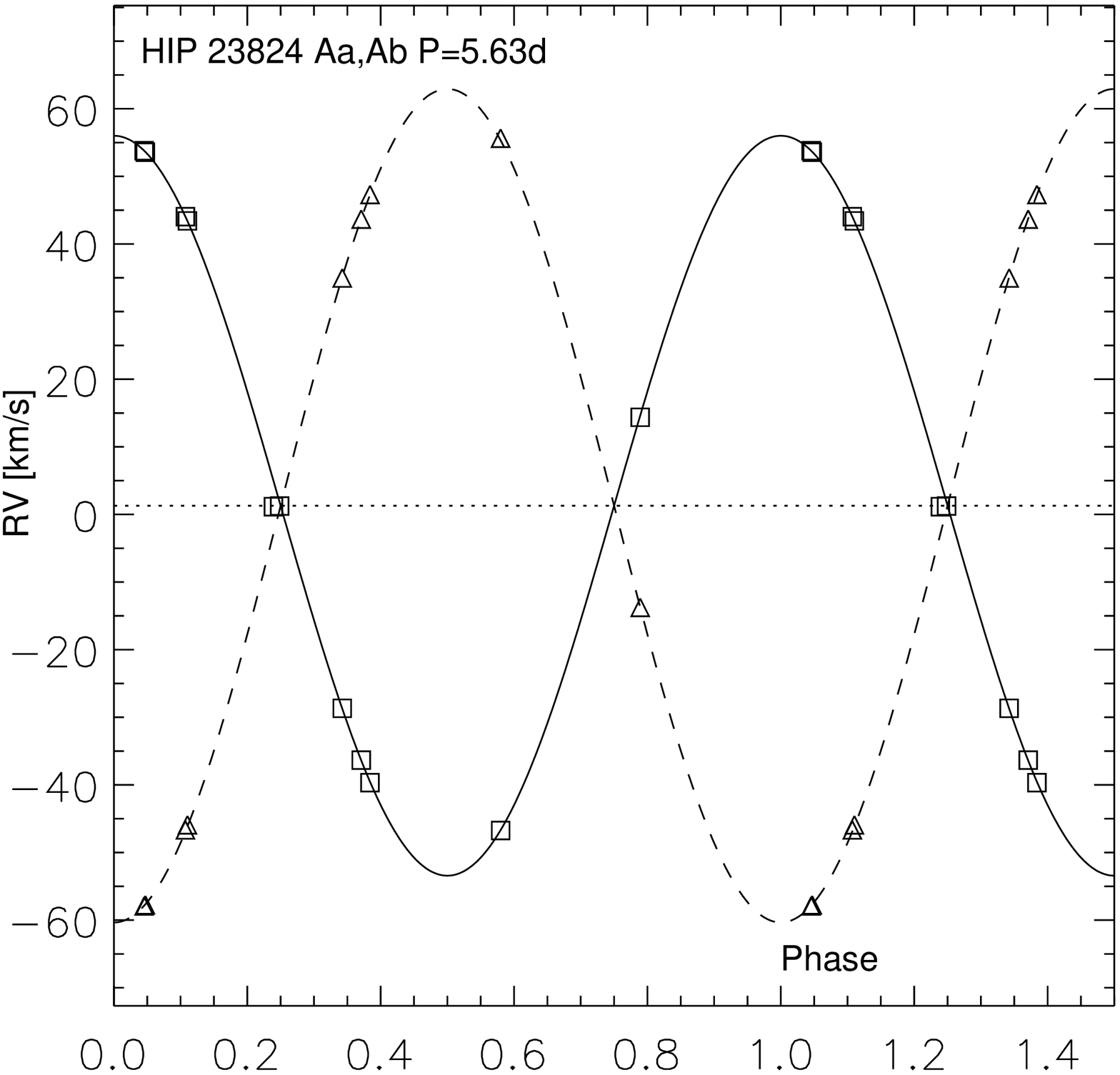}
\caption{Spectroscopic orbit of HIP 23824 Aa,Ab, $P=5.63$\,d.
\label{fig:23824a}
}
\end{figure}

\begin{figure}
\plotone{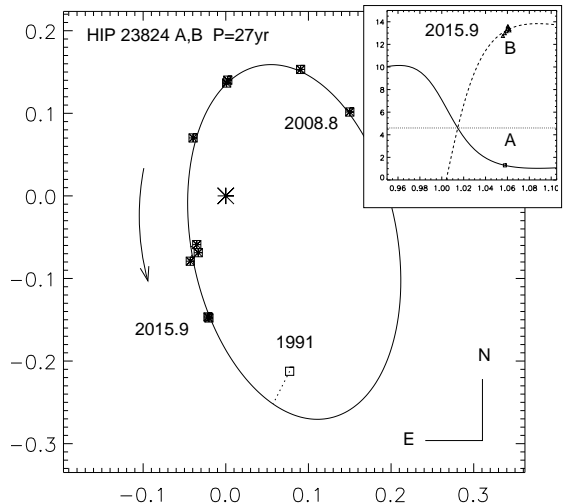}
\caption{Orbit  of  HIP 23824  A,B  (HDS  669),  $P=27.0$\,yr, scale  in
  arcseconds. The insert shows fragment of the RV curve.
\label{fig:HDS669}
}
\end{figure}

Double  lines were  noted by  \citet{N04}, suggesting  a spectroscopic
subsystem.  The orbit  of the subsystem Aa,Ab (Figure~\ref{fig:23824a})
is determined  from several  triple-lined CHIRON spectra.   A circular
orbit is forced, considering the  short period;  I did not subtract the
outer orbit here.

The visual orbit of  the outer system A,B (Figure~\ref{fig:HDS669}) is
only slightly  revised by  adding three  recent measures  to the
latest  published  solution;  all   measures  except  the  first  {\it
  Hipparcos}  resolution come from  SOAR.  The  spectroscopic elements
are tentative  owing to  the lack of  the RV coverage.   No additional
useful RV data were found in the literature or in archives. I used the
$\gamma$-velocity of  Aa,Ab as  a single measure  (the CHIRON  data span
  only 1.15 years), while the individual velocities of B show a small
positive  trend  in  agreement  with  the orbit.   The  data  are  not
sufficient  for  determination  of   the  RV  amplitudes  and  masses.
Instead,  I fixed  $\gamma_{\rm  AB} =  4.6$\,km~s$^{-1}$  to get  the
expected mass ratio $q_{\rm AB} = 0.38$. Unfortunately, the periastron
passage in 2014.3 has not  been covered by spectroscopy, otherwise the
masses could have been measured.

The HIP2 parallax  of  20.8$\pm$0.5\,mas leads to the mass sum of
3.15 ${\cal  M}_\odot$, while  the RV amplitudes  in the orbit  of A,B
correspond to the mass sum of 2.8 ${\cal M}_\odot$. In the following I
adopt the mass sum of 3.0 ${\cal M}_\odot$, or a dynamical parallax of
21.2\,mas.  The HIP2 parallax is consistent within its error; it could
be biased by the orbital motion.

The  system model  is adjusted  to the  data in  the same  way  as for
HIP~13498.  Starting with the mass sum of 3.0 ${\cal M}_\odot$ and the
well-measured mass ratio  of 0.888 in the inner orbit,  I fix the mass
ratio  in  the  outer  orbit  at  $q_{\rm  AB}=0.38$  and  obtain  the
component's  masses  listed  in  Table~\ref{tab:23824}.   The  modeled
combined color $V-K_s  = 1.40$ mag equals the  measured one; the model
predicts  $B-V =  0.55$ mag,  slightly bluer  than  measured.  The
adopted masses  imply an inner  orbital inclination of  48\fdg6, while
the inclination of the outer orbit is 62\degr.

The  lithium line is  detected in  Aa and  Ab, while  it is  below the
detection threshold in  B, just as in HIP  13498.  The parameters of
the   Gaussian   fits   to    the   lithium   line   are   listed   in
Table~\ref{tab:23824}.

\begin{deluxetable}{l c c c }    
\tabletypesize{\scriptsize}     
\tablecaption{Components  of  HIP 23824
\label{tab:23824}          }
\tablewidth{0pt}                                   
\tablehead{                                                                     
\colhead{Parameter} & 
\colhead{Aa} & 
\colhead{Ab} & 
\colhead{B} 
}
\startdata
Mass$^*$ (${\cal M}_\odot$)   & 1.15 & 1.02 & 0.83 \\
$M_V$ (mag)      & 3.89  & 4.72  & 6.20 \\
$T_e^*$ (K)       & 6227 & 5766 & 5111 \\
$(B-V)^*$ (mag)     & 0.48 & 0.69 & 1.02 \\ 
$a_{\rm Li}$ & 0.119    & 0.044  & $<$0.02 \\
              & $\pm$0.005 &  $\pm$0.005 &   \\
$\sigma_{\rm Li}$ (km~s$^{-1}$) & 7.50 & 8.24 & \ldots \\
                   & $\pm$0.36 &  $\pm$1.20 &  \\
$EW_{\rm Li}$ (m\AA)       & 49 & 20 & $<$8 \\
$f_R^*$      &  0.57 & 0.33 & 0.10 \\ 
$EW_{\rm  corr}$ (m\AA)  & 86  & 62 & $<$80 
\enddata 
\end{deluxetable}

The   spatial    velocity   of   HIP~23824    calculated   using   the
$\gamma$-velocity    of   A,B   is    $(U,W,V)   =    (-29.4,   -18.5,
-4.2)$\,km~s$^{-1}$.  A correction of $(-5.8, -4.0)$~mas~yr$^{-1}$ was
applied to the HIP2 proper  motion to subtract the photo-center motion
in  the visual  orbit.   \citet{Eggen1985} believes  that this  triple
system belongs to  the Hyades moving group, although  the agreement of
its velocity is not as good as for HIP~13498.

\section{HIP 113597 (HD 217379)}
\label{sec:113597}

This  K6.5Vke star  is a  known 1\farcs9  visual binary  RST~1154 (WDS
23005$-$2619AB) and an X-ray source.  A brighter star HIP~113579
(HD~217373) at  $581''$ distance on  the sky (projected  separation 18
kAU) forms  with HIP~113597  a co-moving wide  pair, and  both objects
belong  to  the  AB  Doradus   group  of  young  stars  in  the  solar
neighborhood \citep{Zuckerman2004}.  The very faint companion found by
\citet{Cvn2010}  at  5\farcs4  from  HIP~113579  remains  unconfirmed,
possibly not real.  The WDS denotes  the brighter star as C, while the
visual  components of  RST~1154  are denoted  as  A and  B.  To  avoid
further confusion, I keep here  the current WDS designations C, A, and
B  in  decreasing  brightness  order; the  unresolved  photometry  and
spectroscopy of HIP~113597 refers to  AB \citep[C, A, B were designated
  hierarchically     as     A,     Ba,    and     Bb     in][]{FG67a}.
Table~\ref{tab:objects}  lists  basic   properties  of  this  multiple
system.   The  photometry  of  AB  and  its  spectral  type  are  from
\citet{Weis1993}.   As the  parallaxes of  AB and  C are  equal within
errors, in the following I adopt the average parallax of 32.6 mas.

High-resolution  spectroscopy revealed  HIP~113597  as a  triple-lined
system.  This was  discovered independently by \citet{Elliott2014} and
\citet{survey}; both works  are based on the echelle  spectra taken in
2010   with   VLT/UVES    and   1.5-m/FECH,   respectively.    Earlier
spectroscopic  studies did  not recognize  the triple-lined  nature of
this object.  For example, \citet{Malo2014} provide RVs of C and AB as
$7.3\pm0.3$  and $6.6\pm0.3$  km~s$^{-1}$,  respectively, from  single
spectra  taken with  Phoenix  at Gemini-South.   \citet{Zuckerman2004}
detected a strong lithium 6707.8\AA ~line in the component C and noted
its  absence  in  AB  (this  is  confirmed  in  the  CHIRON  spectra).
\citet{N04}  measured the  constant RV(C)=6.10\,km~s$^{-1}$  over 2135
days; they did not observe AB.

The visual  binary A,B was last  measured in 2015.73  with the speckle
camera at  SOAR, yielding the position of  239\fdg3, 2\farcs30 and
the  magnitude  difference  $\Delta  I  =  0.79$  mag.   Its  previous
measurement by {\it Hipparcos}  in 1991.25 is 235\fdg7, 1\farcs839
and $\Delta Hp  = 0.79$ mag.  All three components Aa,  Ab, and B have
comparable  brightness   and  color,  so  $\Delta   m$  is  apparently
independent of the wavelength.

\begin{figure}
\plotone{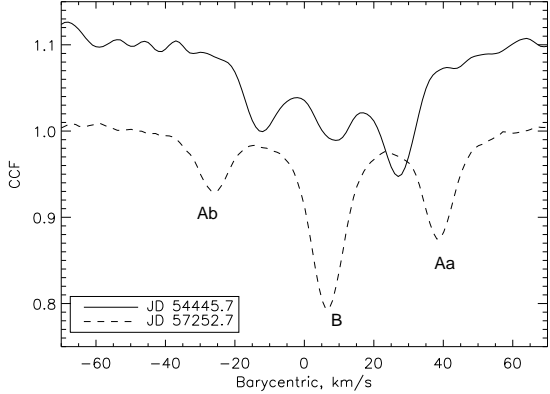}
\caption{CCFs of  HIP 113597 in  two epochs. The upper  curve computed
  from the 2010 FECH spectrum (displaced vertically by 0.1) shows that
  the central  component belonging to B  was weaker than  Aa, while in
  most spectra taken in 2015 (lower curve) the central dip is much stronger.
\label{fig:113597CCF}
}
\end{figure}

\begin{figure}
\plotone{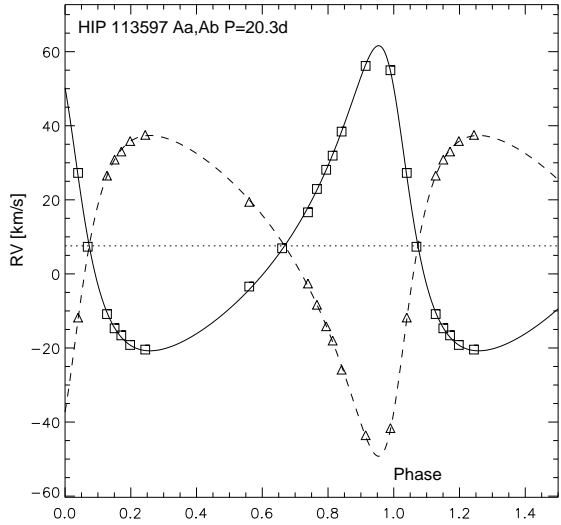}
\caption{Spectroscopic orbit of the inner subsystem HIP 113597 Aa,Ab, $P=20.3$\,d.
\label{fig:113597a}
}
\end{figure}

\begin{figure}
\plotone{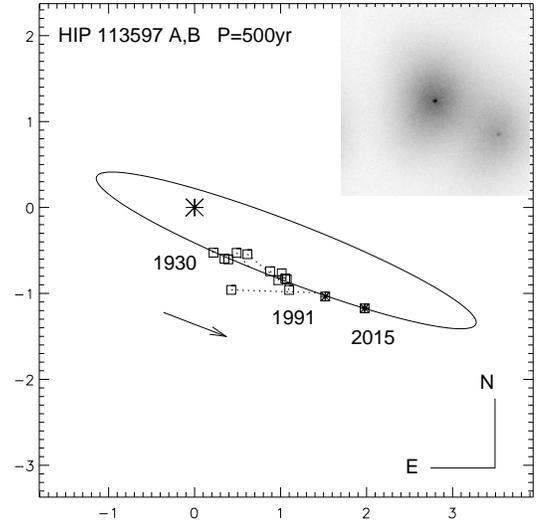}
\caption{Tentative orbit of the pair HIP 113597 A,B (RST 1154). Visual
  measurements are  plotted as empty squares, the  {\it Hipparcos} and
  speckle   data   as   filled   squares.   The   insert   shows   the
  ``shift-and-add'' image in the $I$ band obtained in 2015.7 at SOAR.
\label{fig:113597AB}
}
\end{figure}

All  three dips  in the  CCF are  narrow (Figure~\ref{fig:113597CCF}).
The central  stationary component belongs  to the visual  secondary B,
the two moving components belong to the spectroscopic subsystem Aa,Ab.
The  relative  amplitudes of  the  dips  are  variable.  This  happens
because the 2\farcs3  separation between A and B  is comparable to the
2\farcs7 diameter of the CHIRON  fiber projected on the sky. Depending
on centering, guiding, and seeing, the fraction of the light from each
component that  enters the  fiber is variable.   In the  spectra taken
with FECH in 2010,  the CCF area (EW) of B is half  of the sum of Aa
and Ab, as expected from  the relative photometry of the visual binary
A,B.   In  most  spectra taken  in  2015  with  CHIRON, the  EW(B)  is
comparable to  the sum EW(Aa)+EW(Ab).  Both instruments have  the same
acquisition and  guiding unit; the difference  in relative intensities
could be  caused by a different  offset of the object  position on the
fiber.

To  determine  relative  fluxes in  the  $V$  band,  I use  the  ratio
EW(Aa)/EW(Ab)=0.68$\pm$0.02,  measured reliably,  and  assume that  it
reflects the light ratio (the  CCF area does not depend  on effective
temperature  at $T_e \sim 4000$\,K). Assuming  $\Delta V_{\rm  AB} =  0.79$ mag
from {\it Hipparcos} photometry, I computed the apparent magnitudes of
B, Aa, and Ab listed in Table~\ref{tab:CCF}.

Spectroscopic   orbit   of  the   inner   subsystem   is  plotted   in
Figure~\ref{fig:113597a}.  The  mass ratio of the inner  pair is 0.95,
the  spectroscopic  masses $M  \sin^3  i$  are  0.48 and  0.45  ${\cal
  M}_\odot$.   I used the  RVs measured  by \citet{Elliott2014}  on JD
2455341.4055 with a correction of $+6.15$\,km~s$^{-1}$. The correction
was derived by computing the center-of-mass velocity $V_{\rm A} = (V_1 K_2 +
V_2  K_1)/(K_1  +  K_2)  =  1.41$  km~s$^{-1}$;  it  must  equal  7.56
km~s$^{-1}$  according to my  orbit.  An  RV bias  is expected  when a
semi-resolved  (blended)  visual  binary   is  observed  with  a  slit
spectrograph like UVES,  as guiding is made on  the combined light and
the individual components are no longer centered in the slit.

\begin{deluxetable}{l c c c }    
\tabletypesize{\scriptsize}     
\tablecaption{Components  of  HIP 113597
\label{tab:113597}          }
\tablewidth{0pt}                                   
\tablehead{                                                                     
\colhead{Parameter} & 
\colhead{Aa} & 
\colhead{Ab} & 
\colhead{B} 
}
\startdata
Mass$^*$ (${\cal M}_\odot$)   & 0.61 & 0.57 & 0.60 \\
$M_V$ (mag)       &  8.29 & 8.71 & 8.41 \\
$T_e^*$ (K)       & 4133 & 3992 & 4090 
\enddata 
\end{deluxetable}

Knowing the absolute visual magnitudes of the components
(distance modulus 2.43 mag), I estimated their masses by assuming that
they are  on the main sequence (the  bright star C is  actually on the
main sequence).   The result is given  in Table~\ref{tab:113597}.  The
combined magnitudes of AB in the $J$ and $K_s$ bands derived from this
model are 7.17 and 6.33 mag, respectively.  They compare well with the
2MASS magnitudes of 7.05 and 6.27.  The model leads to the $B-V$ color
of 1.39  mag, while the  measured color is 1.34  mag \citep{Weis1993}.
Good agreement between modeled and actual colors shows that the system
indeed  consists  of three  main-sequence  stars  of $\sim$0.6  ${\cal
  M}_\odot$ each.   The model implies effective  temperatures given in
the  last  line of  Table~\ref{tab:113597}.   The  mass  ratio in  the
subsystem Aa,Ab is 0.95 according to the model, matching the measured ratio.
The orbital inclination of Aa,Ab is $\sim$70\degr. 

The  pair  A,B has  opened  up from  0\farcs6  to  2\farcs3 since  its
discovery by  \citet{Rst1955} in  1930.  We know  the distance  to the
system and its estimated mass sum, 1.8 ${\cal M}_\odot$.  The observed
motion can  be represented  by a visual  orbit matching this  mass sum
(Figure~\ref{fig:113597AB}).   This   orbit  is  {\em   not}  uniquely
constrained by the  data, so the errors of its  elements are undetermined.
It is  computed only to  show that the  motion is compatible  with a
Keplerian orbit and  to get an idea of the period.   Owing to the long
period, no  substantial improvement of  this orbit is expected  in the
following decades.

The  average RV(B)  is 7.05  km~s$^{-1}$ with  a rms  scatter  of 0.54
km~s$^{-1}$  (see Table~\ref{tab:rv}).  The two  measurements  in 2010
average at  8.0 km~s$^{-1}$,  while the 13  measurements in  2015 have
the average  value  of  6.76  km~s$^{-1}$  with  a  rms  scatter  of  0.16
km~s$^{-1}$.  It  is unlikely that  the difference is  attributable to
the different spectrographs, FECH and CHIRON,  as the RVs of Aa and Ab
measured  from  the  same   spectra  do  not  show  instrument-related
systematics.  Motion in the visual orbit is too slow to explain the RV
change   over five   years.  \citet{Elliott2014}   measured  RV(B)=5.5
km~s$^{-1}$ in  2010, with  an uncertain bias  (see above).   I cannot
exclude a low-mass (possibly  planetary) companion to B modulating its
RV.   The outer  orbit  of A,B  leaves  ample room  for such  putative
subsystem.

The positive RV difference  RV(A)$-$RV(B) in 2015 confirms the correct
choice of the ascending node in this provisional orbit.  I measured in
2010   RV(C)=7.03  km~s$^{-1}$   \citep{survey},  in   agreement  with
\citet{N04}.  Analysis of  this  bright component  C  (HIP 113579)  is
outside the scope of this work.

The measured  RV of the component  A together with  the known distance
and  proper motion leads to  the Galactic  velocity of  $(U,V,W) =  (-3.0, -25.5,
-15.1)$ km~s$^{-1}$.  A correction for  the motion in the orbit A,B is
not  applied here;  it  should  be  small, $<$1  km~s$^{-1}$  According  to
\citet{Malo2014},  the AB Doradus  group has  the spatial velocity  $(U,V,W) =
(-7.1, -27.2, -13.8)$ km~s$^{-1}$. 

The   10-yr    photometric   monitoring   of    HIP~113597   by   ASAS
\citep{ASAS}\footnote{http://www.astrouw.edu.pl/asas/}     shows     a
constant brightness of $V=9.58$ mag  with an rms scatter of 0.036 mag;
it is not  identified as a variable star in  the ASAS database.  Young
0.6 ${\cal M}_\odot$  dwarfs are expected to be  active, but no flares
have been detected so far.  The H$\alpha$ line has a low contrast,
  possibly being filled by  emission. The projected rotation velocity
of all three  components is moderate, from 7.2  to 8.8 km~s$^{-1}$. It
is faster than the  estimated pseudo-synchronous rotation in the inner
pair, 3.6 km~s$^{-1}$.

\section{Discussion}
\label{sec:disc}

\subsection{Masses}
\label{sec:mass}

\begin{figure}
\plotone{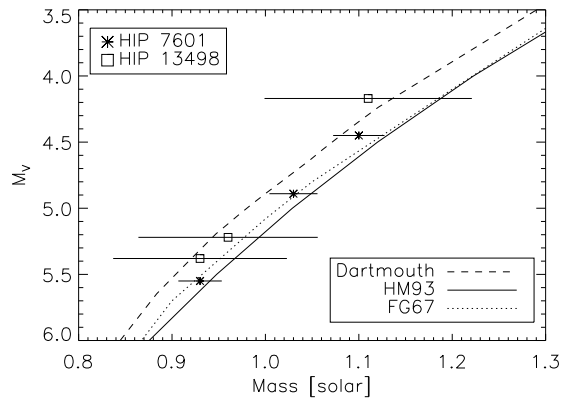}
\caption{Relation between mass and absolute magnitude for HIP 7601 and
  13498. The standard relations are traced by the dashed line \citep[][solar metallicity]{Dotter2008}, 
full line \citep{HM93}, and dotted line \citep{FG67a}.
\label{fig:masslum}
}
\end{figure}

Masses of  the components of HIP  7601 are measured  from the combined
outer orbit  with an accuracy of  2.5\%.  The masses of  HIP 13498 are
constrained by the orbit and parallax only to within $\sim$10\%, while
the  masses  of  HIP  23824  and 113597  are  not  measured  directly.
Figure~\ref{fig:masslum}  plots absolute magnitude  vs.  mass  for the
first  two  systems.   The  data  match empirical  relations  such  as
\citep{HM93,FG67a}.  On the other  hand, the Dartmouth 1-Gyr isochrone
for solar  metallicity is $\sim$0.2  mag brighter; the  discrepancy is
even larger for [Fe/H]=$-0.15$. The agreement with the isochrone would
be   better  for   a  larger-than-solar   metallicity.   Spectroscopic
measurements  of [Fe/H]  could have  been biased  by  the unrecognized
multiplicity.

\subsection{Lithium and rotation}

\begin{figure}
\plotone{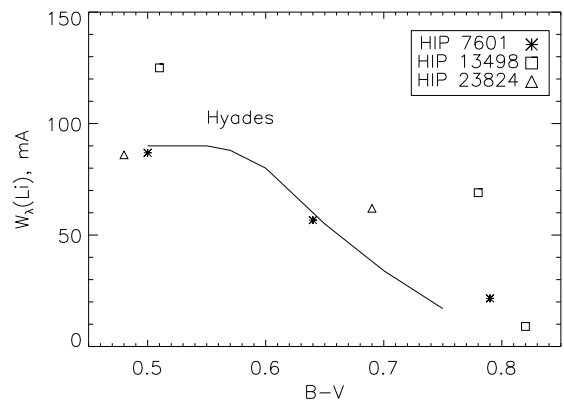}
\caption{Equivalent  width of the  lithium line  in the  components of
  HIP~7601  (squares),  13498  (asterisks), and 23824 (triangles)  vs.   their  $B-V$
  color. The solid line is a schematic trend for the Hyades from Fig.~4
  of \citet{Soderblom1993}.
\label{fig:hyad}
}
\end{figure}

The equivalent  widths of the lithium  line in HIP 7601  and HIP 23824
are  comparable  to   the  stars  of  similar  color   in  the  Hyades
\citep{Soderblom1993},   as   illustrated  in   Figure~\ref{fig:hyad},
suggesting  an  age  of  $\sim$0.6\,Gyr.  However,  large  scatter  of
lithium abundance  and activity indicators among  stars in the Hyades  makes this
estimate rather uncertain.  The lithium  line in HIP 13498 is stronger
than  in the  Hyades, although  the system  may belong  to  this group
kinematically.

The components  of both  HIP~7601 and 13498  have slow  axial rotation
ranging from 1.8 to 4.3  km~s$^{-1}$, as inferred from the width of their
CCFs.   Assuming   that  components   of  the  inner   subsystems  are
synchronized  and  aligned  with  the  orbits,  I  estimate  projected
synchronous velocities of  2.0 and 2.7 km~s$^{-1}$. They  are close to
the actual  $V \sin  i$ of  the secondary components  Ab, 1.8  and 2.8
km~s$^{-1}$  in   HIP  7601  and  13498,   respectively.  The  primary
components Aa and the visual  secondaries B rotate slightly faster. If
the lines  are broadened by  additional factors, compared to  the Sun,
the above estimates are upper  limits.  In the closest inner binary HIP
23824 Aa,Ab,  the synchronous  speed of  Aa, estimated at  $V \sin  i =
7.3$\,km~s$^{-1}$ , matches its measured  rotation of  7.1 km~s$^{-1}$,
while B rotates much slower at $V \sin i = 1.6$\,km~s$^{-1}$.

\subsection{Origin}

By  definition,  triple-lined systems  have  all  three components  of
comparable masses  and luminosities.  If their  components were chosen
randomly from  the stellar mass function, SB3s  would be exceptionally
rare.  The  four nearby SB3s  studied here do not  appear exceptional,
demonstrating  that stars  of  similar masses  are commonly  assembled
together in hierarchical  systems.  It is legitimate to  ask how these
objects  were formed,  given that  the  initial size  of a  solar-mass
protostellar core,   on the order of the  Jeans length ($\sim 10^4$
  AU), is much larger than the size of those multiples. 

It  is currently believed  that binaries  forming by  fragmentation of
cores   or   circumstellar  disks   have   relatively  large   initial
separations, no less than $\sim$50 AU,  owing to the opacity limit
  to  fragmentation   \citep{opacity-limit}.   Subsequent  accretion
shrinks  the binary,  while its  components  grow in  mass.    The
  binary   separation    decreases   roughly   as    ${\cal   M}^{-2}$
  \citep{BB2005}.    The  secondary  grows faster  than the  primary,
naturally leading to close pairs of comparable-mass stars.  To explain
the origin of a hierarchical SB3, we have to admit that the inner pair
formed  first and  became  closer, while  the  outer component  formed
later.  If this triple system later gained a substantial mass and this
mass was accreted preferentially by the distant companion, one expects
that the former  tertiary would become nearly as  massive as the close
pair, converting itself into a primary. Such triple systems indeed are
common  \citep[e.g. $\kappa$~For,][]{Tok13}.  However,  in the  case of
SB3s the growth  of the outer component had to be  regulated in such a
way that all three stars end up with similar masses.  The challenge of
explaining  the formation of  SB3s may  lead to  new insights  on star
formation in general.

All four systems studied here are structured in a similar way, with an
inner spectroscopic pair and a tertiary companion. The orbits of those
tertiaries, however, have quite  different periods ranging from 1.75~yr
to $\sim$500~yrs.  In three  systems the inclinations of  the inner
orbits are close  to the inclinations of the  outer orbits, suggesting
co-planarity.  The inner orbit in  the closest triple, HIP~7601, has a
small  eccentricity  of $e=0.1$.   This  also  argues for  approximate
co-planarity in this system,  because otherwise the Kozai-Lidov cycles
\citep{KCTF} would have increased the inner eccentricity.

On  the other  hand, the  inner  orbits in  HIP 13498  and 113597  are
moderately  eccentric. The  separations at  periastron in  these inner
binaries correspond to equivalent circular orbits with $P_{\rm circ} =
P (1 - e)^{3/2}$ of 7.8  and 8.7 days, respectively. These periods are
shorter  than the  10-day circularization  period of  solar-type stars
\citep{Mayor2000},  meaning that  the components  tidally  interact at
periastron.   The Kozai-Lidov mechanism  cannot be  ruled out.  In any
case, the inner orbits  will eventually become circular.  These triple
systems  are relatively  young  and still  evolve dynamically  towards
circular  inner  orbits. The  inner  binary  in  HIP~23824 is  already
circularized.

The  triple system  HIP~7601 holds  the record  of the  shortest outer
period (639 days) among the 4847 F- and G-type dwarfs within 67\,pc of
the Sun \citep{FG67b}.  The outer periods of multiple  systems in this
sample are  all longer than  $\sim$1000 days. This means  that shorter
outer periods  are rare.  Nevertheless, such  systems certainly exist.
For example, VW~LMi contains four solar-type stars packed in the outer
orbit  with a period  of only  355 days  \citep{Pribulla2008}. Several
triple  stars  with  short   outer  periods  were  recently  found  by
\citet{Borkovits2016}  in   the  {\it  Kepler}   sample  of  eclipsing
binaries.

If  the orbit  of the  tertiary component  shrinks, at  some  point it
starts  to interact  dynamically with  the inner  pair, even  when the
orbits are co-planar (no  Kozai-Lidov cycles). Such interaction should
start at moderate period ratios  (on the order of a few tens) and may
take the form of mean  motion resonances (integer period ratios).  The
period ratio  in HIP~7601  is close (although  not quite equal)  to an
integer  number.   The ratio  of  inner  periods  in the  hierarchical
quadruple   system   HD~91962  is   also   an   integer  number   19.0
\citep{Planetary}. The  absence of the dust disk  and outer companions
to HIP~7601 imply that its outer orbit is no longer shrinking.



\section{Summary}

Orbital elements  of four  hierarchical systems have  been determined
here,  based  on the original  RVs and  complemented by  published  and  new
resolved  measurements. All    objects are  relatively young,  as
inferred from  the chromospheric activity (X-ray  sources), presence of
lithium,  and kinematics.  Measurements  of masses,  luminosities, and
rotation  give an  interesting material  for calibrating  evolution of
solar-type stars and their ages.

The  system  HIP~7601 is  most  interesting,  being  the closest,  the
brightest, and with accurately determined outer orbit and masses.  The
accuracy of  masses can be  further improved by accumulating  more data
and/or by using better instruments.   The semi-major axis of the inner
19-day subsystem is 6\,mas, making it an easy target for long-baseline
interferometers such as  VLTI.  Only a few observations  are needed to
establish the relative orbit orientation, confirming or refuting their
suggested co-planarity.   New data will  also improve the  accuracy of
the measured masses and period  ratio. 

\acknowledgements

I  thank   the  operators  of   the  1.5-m  telescope   for  executing
observations  of  this  program  and  the  SMARTS  team  at  Yale  for
scheduling and pipeline processing.  This work used the SIMBAD service
operated  by  Centre des  Donn\'ees  Stellaires (Strasbourg,  France),
bibliographic references from  the Astrophysics Data System maintained
by SAO/NASA,  the Washington Double  Star Catalog maintained  at USNO,
and  products  of the  2MASS  survey.  HIP~113597  was discussed  with
P.~Elliott. Data from the ESO archive (program 088C.513 and its
continuation) were used.  Comments by anonymous Referee helped to improve the paper.


{\it Facilities:}  \facility{CTIO:1.5m}, \facility{SOAR} 






\clearpage

\LongTables

\setcounter{table}{3}


\begin{deluxetable*}{r l c rrr rrr l}    
\tabletypesize{\scriptsize}     
\tablecaption{Radial velocities and residuals
\label{tab:rv}          }
\tablewidth{0pt}                                   
\tablehead{                                                                     
\colhead{HIP} & 
\colhead{System} & 
\colhead{Date} & 
\colhead{$V_1$} & 
\colhead{$\sigma_1$} & 
\colhead{(O$-$C)$_1$ } &
\colhead{$V_2$} & 
\colhead{$\sigma_2$} & 
\colhead{(O$-$C)$_2$ } &
\colhead{Ref.\tablenotemark{a}} \\ 
 & & 
\colhead{(JD $-$2400000)} &
\multicolumn{3}{c}{(km s$^{-1}$)}  &
\multicolumn{3}{c}{(km s$^{-1}$)}  &
}
\startdata
7601 & Aa,Ab & 54855.5390 &     20.426 &      0.100 &      0.147 &  -24.180 &      0.200 &     -0.080 & H \\
7601 & Aa,Ab & 54856.5450 &     12.032 &      0.100 &      0.206 &   \ldots &      \ldots&     \ldots & H \\  
7601 & Aa,Ab & 54859.5670 &    -17.230 &      9.900 &      0.305 &   20.933 &      0.200 &      0.094 & H \\
7601 & Aa,Ab & 55563.5520 &    -14.128 &      0.100 &      0.090 &   16.785 &      0.200 &     -0.112 & H\\
7601 & Aa,Ab & 55563.6090 &    -13.323 &      0.100 &      0.111 &   \ldots &      \ldots&     \ldots & H \\
7601 & Aa,Ab & 55978.4900 &     24.451 &      0.100 &     -0.074 &  -29.116 &      0.200 &      0.031 & H\\
7601 & Aa,Ab & 55978.4950 &     24.405 &      0.100 &     -0.086 &  -29.264 &      0.200 &     -0.158 & H\\
7601 & Aa,Ab & 56269.7090 &     19.545 &      0.100 &     -0.215 &  -23.685 &      0.200 &     -0.201 & C\\
7601 & Aa,Ab & 56896.8140 &    -38.118 &      0.100 &     -0.052 &   45.101 &      0.150 &     -0.138 & C\\
7601 & Aa,Ab & 56987.5270 &     10.661 &      0.100 &      0.115 &  -12.509 &      0.150 &      0.024 & C\\
7601 & Aa,Ab & 56987.5280 &     10.621 &      0.100 &      0.084 &  -12.517 &      0.150 &      0.005 & C\\
7601 & Aa,Ab & 56991.5900 &    -27.930 &      0.100 &      0.017 &   33.067 &      2.000 &     -0.147 & C\\
7601 & Aa,Ab & 56994.5770 &    -37.149 &      0.100 &     -0.179 &   44.112 &      0.150 &      0.175 & C\\
7601 & Aa,Ab & 57175.9430 &     33.623 &      0.100 &      0.084 &  -40.053 &      0.100 &     -0.193 & C\\
7601 & Aa,Ab & 57226.7530 &    -37.633 &      0.100 &      0.111 &   44.691 &      0.150 &     -0.166 & C\\
7601 & Aa,Ab & 57284.8840 &    -37.582 &      0.100 &      0.126 &   44.774 &      0.150 &     -0.040 & C\\
7601 & Aa,Ab & 57352.5730 &     31.592 &      0.100 &     -0.004 &  -37.246 &      0.150 &      0.304 & C\\
7601 & Aa,Ab & 57365.5720 &     -7.403 &      0.100 &     -0.087 &    8.838 &      0.150 &      0.144 & C\\
7601 & Aa,Ab & 57402.5990 &    -28.871 &      0.100 &     -0.028 &   34.134 &      0.150 &     -0.145 & C\\
 7601 & A,B &  51845.2470 &    -12.060 &      2.000 &     -0.009 &    18.700 &      1.000 &     -0.110 & W \\
 7601 & A,B &  54782.5586 &      6.040 &      1.000 &      0.580 &   -15.950 &      1.000 &     -0.308 & L \\
 7601 & A,B &  54855.5386 &      2.225 &      0.080 &      0.171 &    -9.147 &      0.200 &     -0.207 & H \\
 7601 & A,B &  54859.5670 &      1.715 &      0.080 &     -0.034 &    -7.980 &      0.200 &      0.362 & H \\
 7601 & A,B &  55563.5516 &     -4.915 &      0.080 &     -0.107 &     4.904 &      9.900 &      0.345 & H \\
 7601 & A,B &  55978.4901 &      5.445 &      0.080 &     -0.130 &   -15.558 &      0.200 &      0.310 & H \\
 7601 & A,B &  55978.4951 &      5.478 &      0.080 &     -0.097 &   -15.797 &      0.200 &      0.071 & H \\
 7601 & A,B &  56269.7094 &    -11.455 &      0.080 &     -0.009 &    17.635 &      0.200 &      0.018 & H \\
 7601 & A,B &  56896.8139 &    -10.712 &      0.080 &     -0.041 &    16.253 &      0.150 &      0.159 & C \\
 7601 & A,B &  56987.5272 &    -10.391 &      0.080 &      0.034 &    15.600 &      0.150 &     -0.010 & C \\
 7601 & A,B &  56987.5280 &    -10.365 &      0.080 &      0.060 &    15.579 &      0.150 &     -0.031 & C \\
 7601 & A,B &  56991.5905 &    -10.168 &      0.080 &     -0.021 &    15.056 &      0.150 &     -0.006 & C \\
 7601 & A,B &  56994.5775 &     -9.903 &      0.080 &      0.032 &    14.519 &      0.150 &     -0.126 & C \\
 7601 & A,B &  57175.9426 &      2.928 &      0.080 &     -0.087 &   -10.800 &      0.150 &      0.033 & C \\
 7601 & A,B &  57226.7528 &      4.950 &      0.080 &      0.034 &   -14.626 &      0.150 &     -0.054 & C \\
 7601 & A,B &  57284.8841 &      5.990 &      0.080 &      0.100 &   -16.544 &      0.150 &     -0.055 & C \\
 7601 & A,B &  57352.5726 &      5.147 &      0.080 &      0.096 &   -14.870 &      0.150 &     -0.032 & C \\
 7601 & A,B &  57365.5723 &      4.613 &      0.080 &      0.028 &   -14.057 &      0.150 &     -0.136 & C \\
 7601 & A,B &  57402.5989 &      2.558 &      0.080 &     -0.043 &   -10.026 &      0.150 &     -0.009 & C \\
13498  &  Aa,Ab &54781.7194 &    -28.948 &      0.500 &     -0.009 &   33.452 &      0.500 &     -0.119 & C\\
13498  &  Aa,Ab &56910.8509 &    -26.499 &      0.050 &      0.012 &   30.725 &      0.050 &     -0.030 & C\\
13498  &  Aa,Ab &56938.8271 &    -27.214 &      0.050 &     -0.060 &   31.456 &      0.050 &     -0.044 & C\\
13498  &  Aa,Ab &57003.7262 &     17.868 &      0.050 &      0.044 &  -20.594 &      0.050 &      0.083 & C\\
13498  &  Aa,Ab &57008.7007 &    -28.481 &      0.050 &     -0.031 &   32.983 &      0.050 &     -0.021 & C\\
13498  &  Aa,Ab &57010.7707 &    -20.229 &      0.050 &      0.007 &   23.464 &      0.050 &     -0.010 & C\\
13498  &  Aa,Ab &57015.6590 &     51.560 &      0.050 &     -0.051 &  -59.917 &      0.050 &     -0.045 & C\\
13498  &  Aa,Ab &57319.6708 &    -25.755 &      0.050 &      0.047 &   29.915 &      0.050 &     -0.017 & C\\
13498  &  Aa,Ab &57333.6875 &    -26.047 &      0.050 &      0.031 &   30.299 &      0.050 &      0.048 & C\\
13498 & A,B  & 51847.307  &     19.650 &      5.000 &      0.306 &   16.700 &      5.000 &     -0.079 & W\\
13498 & A,B  & 54781.7194 &     20.430 &      0.500 &      0.006 &   14.340 &      0.500 &      0.257 & L\\
13498 & A,B  & 56910.8510 &     17.816 &      0.040 &      0.004 &   20.613 &      0.040 &      0.010 & C\\
13498 & A,B  & 56938.8270 &     17.762 &      0.040 &     -0.043 &   20.581 &      0.040 &     -0.038 & C\\
13498 & A,B  & 57003.7260 &     17.853 &      0.040 &      0.063 &   20.822 &      0.150 &      0.167 & C\\
13498 & A,B  & 57008.7010 &     17.769 &      0.040 &     -0.020 &   20.650 &      0.040 &     -0.008 & C\\
13498 & A,B  & 57010.7710 &     17.792 &      0.040 &      0.003 &   20.646 &      0.040 &     -0.013 & C\\
13498 & A,B  & 57015.6590 &     17.736 &      0.040 &     -0.051 &   20.712 &      0.040 &      0.050 & C\\
13498 & A,B  & 57319.6710 &     17.737 &      0.040 &      0.007 &   20.852 &      0.040 &      0.048 & C\\
13498 & A,B  & 57333.6880 &     17.755 &      0.040 &      0.027 &   20.793 &      0.040 &     -0.017 & C\\
23824 & Aa,Ab & 57258.8734 &     14.375 &      0.500 &     -0.186 &    -13.756 &      0.500 &     -0.100 & C\\
23824 & Aa,Ab & 57333.8850 &     43.417 &      0.500 &      0.013 &    -45.911 &      0.500 &      0.237 & C\\
23824 & Aa,Ab & 57350.7610 &     44.042 &      0.500 &     -0.015 &    -46.657 &      0.500 &      0.227 & C\\
23824 & Aa,Ab & 57364.6842 &    -46.749 &      0.500 &     -0.062 &     55.650 &      0.500 &      0.308 & C\\
23824 & Aa,Ab & 57374.6089 &    -28.670 &      0.500 &      0.029 &     34.973 &      0.500 &     -0.105 & C\\
23824 & Aa,Ab & 57378.5676 &     53.788 &      0.500 &     -0.014 &    -57.912 &      0.500 &     -0.051 & C\\
23824 & Aa,Ab & 57378.5777 &     53.567 &      0.500 &     -0.058 &    -57.742 &      0.500 &     -0.080 & C\\
23824 & Aa,Ab & 57379.6591 &      1.240 &      2.500 &     -0.564 &      1.150 &      9.500 &     -3.856 & C\\
23824 & Aa,Ab & 57391.6627 &    -36.338 &      0.500 &      0.049 &     43.675 &      0.500 &     -0.064 & C\\
23824 & Aa,Ab & 57391.7363 &    -39.662 &      0.500 &     -0.147 &     47.314 &      0.500 &      0.052 & C\\
23824 & A,B &  57333.8850 &      \ldots& \ldots    & \ldots   &    12.751 &      0.200 &     -0.301 & C\\
23824 & A,B &  57350.7610 &      \ldots& \ldots    & \ldots   &    12.989 &      0.200 &     -0.167 & C\\
23824 & A,B &  57364.6842 &      1.290 &  0.200    &   -0.003 &    13.166 &      0.200 &     -0.068 & C\\
23824 & A,B &  57374.6089 &      \ldots& \ldots    & \ldots   &    13.406 &      0.200 &      0.121 & C\\
23824 & A,B &  57378.5676 &      \ldots& \ldots    & \ldots   &    13.622 &      0.200 &      0.317 & C\\
23824 & A,B &  57378.5777 &      \ldots& \ldots    & \ldots   &    13.558 &      0.200 &      0.253 & C\\
23824 & A,B &  57391.6627 &      \ldots& \ldots    & \ldots   &    13.238 &      0.200 &     -0.128 & C\\
23824 & A,B &  57391.7363 &      \ldots& \ldots    & \ldots   &    13.350 &      0.200 &     -0.017 & C\\
113597  & Aa,Ab  &55341.4035 &     56.140 &      1.000 &     -0.228 &     -43.560 &      1.000 &      0.730 & E \\
113597  & Aa,Ab  &55444.6827 &     54.973 &      0.500 &      0.312 &     -41.635 &      0.500 &      0.841 & F \\
113597  & Aa,Ab  &55445.6957 &     27.253 &      0.500 &      1.041 &     -11.788 &      0.500 &      0.467 & F \\
113597  & Aa,Ab  &57218.7357 &    -16.610 &      1.500 &      0.368 &      33.040 &      1.500 &     -0.586 & C \\
113597  & Aa,Ab  &57226.6538 &     -3.420 &      1.500 &      0.796 &      19.440 &      1.500 &     -0.629 & C \\
113597  & Aa,Ab  &57228.6877 &      6.910 &      5.000 &      0.387 &       \ldots&     \ldots &     \ldots & C \\     
113597  & Aa,Ab  &57250.6373 &     16.618 &      0.500 &     -1.144 &      -2.600 &      1.500 &      0.679 & C \\
113597  & Aa,Ab  &57251.7665 &     28.105 &      0.500 &      0.228 &     -14.146 &      0.500 &     -0.122 & C \\
113597  & Aa,Ab  &57252.7201 &     38.411 &      0.500 &      0.177 &     -25.867 &      0.500 &     -0.841 & C \\
113597  & Aa,Ab  &57277.7028 &      7.325 &      5.200 &     -2.327 &       \ldots&     \ldots &     \ldots & C \\
113597  & Aa,Ab  &57299.7150 &    -14.700 &      1.000 &     -0.219 &      30.850 &      0.500 &     -0.123 & C \\
113597  & Aa,Ab  &57300.6760 &    -19.200 &      0.500 &     -0.144 &      35.870 &      0.500 &      0.037 & C \\
113597  & Aa,Ab  &57301.6130 &    -20.440 &      0.500 &      0.071 &      37.520 &      0.500 &      0.142 & C \\
113597  & Aa,Ab  &57319.5924 &    -10.857 &      0.500 &     -0.504 &      26.531 &      0.500 &     -0.056 & C \\
113597  & Aa,Ab  &57332.5887 &     22.925 &      0.500 &      0.456 &      -8.337 &      0.500 &     -0.058 & C \\
113597  & Aa,Ab  &57333.5621 &     31.921 &      0.500 &     -0.083 &     -18.005 &      0.500 &      0.403 & C \\
113597  & Aa,Ab  &57364.5277 &    -18.394 &      0.500 &      0.596 &      35.434 &      0.500 &     -0.329 & C \\
113597  & Aa,Ab  &57377.5298 &     58.780 &      0.500 &     -0.573 &     -48.380 &      0.500 &     -0.919 & C \\
113597  & B      &  55444.6827 &   7.719 &  \ldots &  \ldots &  \ldots &  \ldots & \ldots & F \\
113597  & B      &  55445.6957 &   8.290 &  \ldots &  \ldots &  \ldots &  \ldots & \ldots & F \\
113597  & B      &  57218.7357 &   6.517 &  \ldots &  \ldots &  \ldots &  \ldots & \ldots & C \\
113597  & B      &  57226.6538 &   7.070 &  \ldots &  \ldots &  \ldots &  \ldots & \ldots & C \\
113597  & B      &  57228.6877 &   6.912 &  \ldots &  \ldots &  \ldots &  \ldots & \ldots & C \\
113597  & B      &  57251.7665 &   6.857 &  \ldots &  \ldots &  \ldots &  \ldots & \ldots & C \\
113597  & B      &  57252.7201 &   6.752 &  \ldots &  \ldots &  \ldots &  \ldots & \ldots & C \\
113597  & B      &  57299.7157 &   6.542 &  \ldots &  \ldots &  \ldots &  \ldots & \ldots & C \\
113597  & B      &  57300.6762 &   6.803 &  \ldots &  \ldots &  \ldots &  \ldots & \ldots & C \\
113597  & B      &  57301.6131 &   6.722 &  \ldots &  \ldots &  \ldots &  \ldots & \ldots & C \\
113597  & B      &  57319.5924 &   6.661 &  \ldots &  \ldots &  \ldots &  \ldots & \ldots & C \\
113597  & B      &  57332.5887 &   6.979 &  \ldots &  \ldots &  \ldots &  \ldots & \ldots & C \\
113597  & B      &  57333.5621 &   6.746 &  \ldots &  \ldots &  \ldots &  \ldots & \ldots & C \\
113597  & B      &  57364.5277 &   6.608 &  \ldots &  \ldots &  \ldots &  \ldots & \ldots & C \\
113597  & B      &  57377.5298 &   6.722 &  \ldots &  \ldots &  \ldots &  \ldots & \ldots & C 
\enddata 
\tablenotetext{a}{
C: CHIRON;
E: \citet{Elliott2014};
F: FECH;
H: HARPS; 
L: \citet{survey};
W: \citet{Wichman2003}. 
}
\end{deluxetable*}

\begin{deluxetable*}{r l l rrr rr l}    
\tabletypesize{\scriptsize}     
\tablecaption{Position measurements and residuals
\label{tab:speckle}          }
\tablewidth{0pt}                                   
\tablehead{                                                                     
\colhead{HIP} & 
\colhead{System} & 
\colhead{Date} & 
\colhead{$\theta$} & 
\colhead{$\rho$} & 
\colhead{$\sigma$} & 
\colhead{(O$-$C)$_\theta$ } & 
\colhead{(O$-$C)$_\rho$ } &
\colhead{Ref.\tablenotemark{a}} \\
 & & 
\colhead{(yr)} &
\colhead{(\degr)} &
\colhead{(\arcsec)} &
\colhead{(\arcsec)} &
\colhead{(\degr)} &
\colhead{(\arcsec)} &
}
\startdata
 7601 & A,B &  2011.0366 &     50.2 &   0.0378 &    0.020 &     -3.7 &  -0.0096 &SOAR  \\
 7601 & A,B &  2011.0366 &     53.4 &   0.0488 &    0.010 &     -0.5 &   0.0014 &SOAR  \\
 7601 & A,B &  2014.7661 &    113.0 &   0.0640 &    0.001 &     -3.0 &  -0.0004 &SOAR  \\
 7601 & A,B &  2014.8535 &    132.4 &   0.0630 &    0.001 &     -1.0 &  -0.0013 &SOAR  \\
 7601 & A,B &  2015.0286 &    173.3 &   0.0586 &    0.001 &     -0.3 &   0.0022 &SOAR  \\
 7601 & A,B &  2015.0286 &    173.3 &   0.0574 &    0.001 &     -0.3 &   0.0010 &SOAR  \\
 7601 & A,B &  2015.1051 &    194.8 &   0.0564 &    0.001 &      0.4 &   0.0013 &SOAR  \\
 7601 & A,B &  2015.1713 &    212.0 &   0.0548 &    0.001 &     -0.2 &  -0.0020 &SOAR  \\
 7601 & A,B &  2015.2504 &    231.4 &   0.0620 &    0.001 &      0.3 &   0.0000 &SOAR  \\
 7601 & A,B &  2015.4976 &    270.9 &   0.0826 &    0.001 &      0.4 &  -0.0014 &SOAR  \\
 7601 & A,B &  2015.7377 &    296.1 &   0.0917 &    0.001 &      0.4 &   0.0009 &SOAR  \\
 7601 & A,B &  2015.9102 &    315.7 &   0.0806 &    0.001 &      0.7 &   0.0006 &SOAR  \\
 7601 & A,B &  2016.0475 &    337.0 &   0.0624 &    0.001 &     -0.4 &  -0.0008 &SOAR  \\
 7601 & A,B &  2016.1344 &    360.1 &   0.0524 &    0.001 &      0.6 &   0.0001 &SOAR  \\
13498 & A,B  & 1927.0200 &    126.4 &    0.390 &    1.050 &      0.3 &   -0.125 & Vis \\
13498 & A,B  & 1933.0000 &    134.2 &    0.460 &    0.050 &      2.3 &   -0.043 & Vis \\
13498 & A,B  & 1936.7000 &    134.5 &    0.410 &    0.050 &     -1.3 &   -0.063 & Vis \\
13498 & A,B  & 1938.8400 &    138.1 &    0.420 &    0.050 &     -0.2 &   -0.028 & Vis \\
13498 & A,B  & 1946.7300 &    145.4 &    0.300 &    0.050 &     -7.2 &    0.000 & Vis \\
13498 & A,B  & 1960.8800 &     95.5 &    0.250 &    0.050 &     -2.0 &    0.015 & Vis \\
13498 & A,B  & 1975.7230 &    123.9 &    0.460 &    0.050 &      0.5 &   -0.045 & Vis \\
13498 & A,B  & 1978.7960 &    128.2 &    0.530 &    0.050 &      1.8 &    0.015 & Vis \\
13498 & A,B  & 1989.9330 &    137.7 &    0.452 &    0.005 &     -0.2 &   -0.001 & Vis  \\
13498 & A,B  & 1990.0500 &    137.3 &    0.310 &    1.050 &     -0.7 &   -0.142 & Vis  \\
13498 & A,B  & 1990.9160 &    139.2 &    0.444 &    0.005 &      0.1 &    0.004 & Spe  \\
13498 & A,B  & 1991.2500 &    140.0 &    0.437 &    0.010 &      0.5 &    0.002 & Hip \\
13498 & A,B  & 1991.6000 &    133.4 &    0.450 &    0.050 &     -6.6 &    0.020 & Vis \\
13498 & A,B  & 1991.7130 &    140.0 &    0.427 &    0.005 &     -0.2 &   -0.002 & Spe  \\
13498 & A,B  & 2010.9660 &     88.7 &    0.179 &    0.002 &     -0.4 &    0.000 & SOAR \\
13498 & A,B  & 2014.7690 &    107.6 &    0.316 &    0.005 &      1.9 &    0.003 & SOAR \\
13498 & A,B  & 2015.0290 &    106.2 &    0.320 &    0.002 &     -0.1 &   -0.001 & SOAR \\
23824 & A,B &  1991.2500 &    200.0 &    0.226 &    0.020 &      7.2 &   -0.032 & SOAR\\
23824 & A,B &  2008.7729 &    304.1 &    0.181 &    0.002 &     -0.2 &    0.002 & SOAR\\
23824 & A,B &  2010.9684 &    329.4 &    0.178 &    0.005 &      0.1 &    0.002 & SOAR\\
23824 & A,B &  2013.1283 &    359.6 &    0.137 &    0.003 &      0.6 &   -0.000 & SOAR\\
23824 & A,B &  2013.1283 &    359.0 &    0.140 &    0.002 &      0.0 &    0.003 & SOAR\\
23824 & A,B &  2014.0430 &     29.4 &    0.081 &    0.003 &      0.7 &    0.004 & SOAR\\
23824 & A,B &  2015.0286 &    149.2 &    0.069 &    0.002 &      3.8 &   -0.006 & SOAR\\
23824 & A,B &  2015.0286 &    154.1 &    0.076 &    0.003 &      8.7 &    0.002 & SOAR\\
23824 & A,B &  2015.1051 &    151.5 &    0.090 &    0.007 &      1.5 &    0.008 & SOAR\\
23824 & A,B &  2015.9102 &    172.1 &    0.149 &    0.003 &     -0.9 &   -0.001 & SOAR\\
23824 & A,B &  2015.9102 &    171.7 &    0.148 &    0.002 &     -1.4 &   -0.003 & SOAR\\
113597  &  A,B  & 1930.2000 &    202.7 &    0.570 &    0.050 &      6.0 &    0.070 & Vis  \\
113597  &  A,B  & 1935.8500 &    210.2 &    0.690 &    0.050 &      2.7 &    0.081 & Vis  \\
113597  &  A,B  & 1941.3300 &    213.0 &    0.720 &    0.050 &     -1.6 &   -0.006 &  Vis \\
113597  &  A,B  & 1950.8600 &    222.8 &    0.720 &    0.050 &      0.2 &   -0.221 &  Vis \\
113597  &  A,B  & 1963.8000 &    228.5 &    0.820 &    0.100 &     -0.5 &   -0.417 &  Vis \\
113597  &  A,B  & 1965.8400 &    229.0 &    1.290 &    0.100 &     -0.8 &    0.007 &  Vis \\
113597  &  A,B  & 1966.6200 &    229.8 &    1.150 &    0.100 &     -0.2 &   -0.151 & Vis  \\
113597  &  A,B  & 1966.8200 &    228.9 &    1.460 &    0.100 &     -1.2 &    0.155 & Vis  \\
113597  &  A,B  & 1966.8500 &    232.9 &    1.270 &    0.100 &      2.8 &   -0.036 & Vis  \\
113597  &  A,B  & 1967.4900 &    232.0 &    1.360 &    0.100 &      1.7 &    0.040 & Vis \\
113597  &  A,B  & 1970.5500 &    231.9 &    1.340 &    0.100 &      0.6 &   -0.048 & Vis  \\
113597  &  A,B  & 1986.5700 &    204.0 &    1.050 &    9.500 &    -31.1 &   -0.685 & Vis \\
113597  &  A,B  & 1991.2500 &    235.7 &    1.839 &    0.010 &     -0.3 &    0.008 & Hip \\
113597  &  A,B  & 2015.7380 &    239.3 &    2.300 &    0.005 &      0.0 &   -0.001 & SOAR 
\enddata 
\tablenotetext{a}{
HIP: Hipparcos;
SOAR: speckle interferometry at SOAR;
Spe: speckle interferometry at other telescopes ;
Vis: visual micrometer measures.
}
\end{deluxetable*}

\end{document}